\documentclass[10pt,journal,onecolumn,final]{IEEEtran}
\usepackage{graphicx}
\usepackage{amsmath}
\allowdisplaybreaks
\usepackage{amssymb}
\usepackage{graphics}
\usepackage{caption}
\usepackage{subcaption}
\usepackage[dvips]{epsfig}
\usepackage{color}
\usepackage[english]{babel}
\usepackage{csquotes}
\MakeOuterQuote{"}

\usepackage{tabulary}
\newcolumntype{K}[1]{>{\centering\arraybackslash}m{#1}}
\usepackage{mathtools}
\usepackage{amsmath}
\usepackage{amsfonts}
\usepackage{amssymb}
\usepackage{graphicx}
\selectfont

\begin{document}
\title{A New Stable Peer-to-Peer Protocol with Non-persistent Peers}
\author{Omer~Bilgen,~\IEEEmembership{Member,~IEEE,
} and~Aaron~B.~Wagner,~\IEEEmembership{Senior Member,~IEEE,} 
}

\maketitle
\let\thefootnote\relax\footnote{Shortened version of this manuscript
was presented at IEEE INFOCOM 2017}
\section{Abstract}Recent studies have suggested that the stability of peer-to-peer networks may rely on \textit{persistent peers}, who dwell on the network after they obtain the entire file. In the absence of such peers, one piece becomes extremely rare in the network, which leads to instability. Technological developments, however, are poised to reduce the incidence of persistent peers,
giving rise to a need for a protocol that guarantees stability with non-persistent peers. We propose a novel peer-to-peer protocol, \textit{the group suppression protocol}, to ensure the stability of peer-to-peer networks under the scenario that all the peers adopt non-persistent behavior. Using a suitable Lyapunov potential function, the group suppression protocol is proven to be stable when the file is broken into two pieces, and detailed experiments demonstrate the stability of the protocol for arbitrary number of pieces. We define and simulate a decentralized version of this protocol for practical applications. Straightforward incorporation of the group suppression protocol into BitTorrent while retaining most of BitTorrent's core mechanisms is also presented. Subsequent simulations show that under certain assumptions, BitTorrent with the official protocol cannot escape from the missing piece syndrome, but BitTorrent with group suppression does.

\begin{IEEEkeywords}
peer-to-peer, BitTorrent, stability, file-sharing, content distribution, Lyapunov, stable protocol, missing piece, rarest.
\end{IEEEkeywords}
\section{Introduction}\label{sect:introduction}
In a peer-to-peer network, a file is divided into a number of pieces, and each peer uploads the pieces obtained so far to other peers while it continues to download the remaining pieces. This results in high utilization of the bandwidth of all peers leading to improved scalability over a traditional server-client structure. Such systems have been deployed widely, as evident from their number of users and the fraction of overall internet traffic they command \cite{Cisco2015}, and they are poised to become even more prominent with the rise of fog networking \cite{fogBonomi, fogVaquero}. A high level p2p protocol can be understood to be a set of rules that governs how a peer contacts another peer and chooses the piece to upload/download to/from that peer, depending on whether a "push" or "pull" policy is employed, respectively. \textit{Network} and \textit{system} will be used interchangeably to refer to the collection of all peers throughout the paper. Also we define a peer to be an \textit{incomplete peer} if it does not have all the pieces of the file and a \textit{seed} if it does.

Peer-to-peer networks with random peer and piece selection policies have been shown to struggle with what Hajek and Zhu \cite{HajekMissing} call \textit{the missing piece syndrome} in which very few peers possess a missing piece while most of the peers, referred to as \textit{the one club} in \cite{HajekMissing}, have all the pieces but the missing piece \cite{HajekMissing, Mathieu}. A qualitative argument why such an imbalanced state persists in the network is as follows: Assume that the one club currently dominates the network, uniformly random peer and piece selection policies are in effect, the peers are non-persistent in the sense that the peers depart as soon as they have the entire file, and the new peers arrive into the network with no pieces. As a result of the uniformly random contact policy, any new peer will contact peers mostly from the one club and thereby become a member of the one club before it can obtain the missing piece. Even when the new peers obtain the missing piece before they are pulled to the one club, they would download the rest of the file quickly thanks to the large upload capacity of the one club. Therefore, they would depart without disseminating the missing piece to other peers long enough to significantly decimate the one club. Hence, the number of peers in the system diverges to infinity once the system is overwhelmed by the one club. A more thorough discussion of the dynamics of the missing piece syndrome can be found in~\cite{HajekMissing}. In the same paper, it has been proven that if the upload rate of the fixed seed is greater than the arrival rate of new peers, who carry no pieces under the model, the system is stable. On the other hand, if the arrival rate of peers exceeds the upload rate of the fixed seed, then the system experiences the missing piece syndrome and is unstable. 

With an ideal p2p protocol the system should remain stable irrespective of the relationship between the arrival rate and the upload rate of the seed, as the arrival rate cannot be internally controlled by the protocol. Zhu and Hajek~\cite{HajekWaiting} establish that if peers linger in the system on average long enough to upload one piece, then the system is stable regardless of the arrival rate. This highlights the significance of having persistent peers from a stability point of view. Based on this study, one could reasonably posit that persistent peers are the main reason why many experimental works evaluating actual trace logs of large p2p networks find their performance to be satisfactory.

As per capita media consumption increases~\cite{UCSD:Information} and moves to 
wireless devices~\cite{Cisco2015}, however, one expects the incidence of 
persistent peers to decline, for three reasons. 
First, data consumption on wireless devices is generally priced 
differently than for wired terminals. In particular,
most wireless providers in the U.S. charge users 
or lower users' speed when their data usage exceeds a certain 
threshold, whereas most wired data plans allow for 
unlimited data transfers, constrained only by the rate.
In the wireless case, uploading is counted toward the overall data usage 
as much as downloading is, and this constitutes a substantial incentive 
for a mobile p2p user to behave non-persistently.
Second, power consumption is a significant concern for mobile users, and 
hence it is less likely for a mobile p2p user to dwell in the system 
after the download is complete than it is for a user of a wired system.
Finally, increases in per-user data consumption~\cite{UCSD:Information}, 
in the absence of attending increases in network capacity, will strain
users' access and discourage them from being persistent peers.
One should note that it is difficult for a protocol to penalize
peers for being non-persistent since, by definition, non-persistent
peers have left the p2p network and, therefore, lie outside its province.

There have been noteworthy attempts to devise stable protocols with non-persistent peers and arbitrary arrival rates. Reittu \cite{Reittu} suggests that all the peers sample three random peers and download a piece chosen uniformly at random from the set of pieces that are found in exactly one of these three peers, and they skip if the set is empty. Experimental results appear to demonstrate the stability of the protocol when the system with one fixed seed begins with the extreme case of all other peers lacking only one common piece, although no proof was provided for stability. In a follow-up work, Norros \textit{et al}. \cite{Norros} showed the global stability of this protocol under large deterministic system limits when the file is composed of two pieces. Later, global stability of the underlying stochastic system for arbitrary number of pieces was proved by O\v{g}uz \textit{et al}. in \cite{Venkat}. In the same work, O\v{g}uz \textit{et al}. \cite{Venkat} also provide a new provably stable scheme in which only new peers that arrive with no piece apply the rule in \cite{Reittu} and the peers lacking only one piece contact $m$ peers and download the last piece only if every piece they have shows up at least twice in aggregate piece profile of these $m$ peers. The protocol requires all other peers to contact a randomly chosen peer and download a piece chosen uniformly at random. The implementation of such a protocol burdens the one club peers in the sense that they are not only prevented from downloading the last piece and consequently departing, but also they are required to keep uploading to the network while waiting to obtain the last piece. Thus, concerns for power consumption of mobile devices and internet data usage may make this protocol unappealing to some p2p users. The same concern can be raised for the protocol in~\cite{Reittu}.

All of the above-mentioned protocols either rely on persistent peers
\cite{HajekWaiting} or on the willingness of peers to decline a useful piece
that is offered to them \cite{Reittu,Norros,Venkat}. We introduce a protocol
that requires neither of these assumptions: the peers may be non-persistent,
and they always download a piece when they connect with a peer
who has a piece that they lack; we call peers that exhibit the latter
behavior \emph{opportunistic.} Instead, we modify the behavior
of the transmitting peers:
peers decline to transmit a piece when the transmission of this
piece would harm stability. When all peers have perfect knowledge
of the state of the network, it makes no difference whether the
uploading peer or the downloading peer is charged with preventing undesirable
piece exchanges. When there is imperfect state information, however,
the uploading peer and downloading peer can have different views of the
network and might reach different conclusions about the desirability
of transmitting a particular piece. In such situations, it is
preferable to rely on the uploading peer to decide whether the
piece should be transmitted, due to a fundamental asymmetry
in the roles of the two peers: the downloading peer has more at stake
than the uploading peer does in the transaction. This is especially
true when the downloading peer has all but one piece and the uploading peer
has the piece that the downloading peer lacks. A downloading peer in this situation has a great
incentive to accept the last piece, irrespective of what the
protocol dictates, thereby completing its copy and allowing
it to leave the network before its misbehavior can be detected.

Recognizing that the formation and persistence of the one club lies at the heart of instability as discussed in \cite{HajekMissing}, we introduce the \textit{group 
suppression protocol} that achieves stability by stopping the one club from recruiting new members. This goal is accomplished in the following way: First define a \textit{group} as a collection of peers that share the same piece profile. And a group of peers whose population is strictly greater than any other group of peers in the network is called \textit{the largest club}. 
The group suppression protocol dictates that a peer from the largest club uploads only to peers who hold greater number of pieces than it does and to refuse the upload to all the other peers when it contacts them. 
An upload from a peer in the largest club to any peer holding more pieces is allowed because such an upload does not draw new members into the largest club. To compare the group suppression protocol to the one suggested by O\v{g}uz \textit{et al}. in~\cite{Venkat}, consider a scenario in which the largest club consists of peers with all the pieces but one common piece, which is also the one club. In the former protocol, the largest club members download the last piece and leave as soon as they encounter a peer that has the last piece. 
Furthermore, they upload nothing since there are no peers that are eligible to receive a piece from them. 
In the latter protocol, when the largest club peers encounter the last piece, they may be prohibited from downloading it based on the piece profile of their $m$ contacts, and they continue uploading while waiting for another contact. Also, when the largest club is very crowded, its peers' waiting times are consequently prolonged. The group suppression protocol compensates the largest club for their extended wait time in the network by allowing them to reduce their upload rate, which constitutes another desirable feature of the protocol. 

In a similar spirit to the group suppression protocol, throttling the upload capacity of peers or even servers has been proposed to increase the performance of p2p systems. Zhang \textit{et al}. \cite{ZhangServers} considered a hybrid p2p system with a file consisting of a single piece and established via a fluid model that throttling the server capacity under certain conditions minimizes the number of peers compared with utilizing the entire capacity of the server. Their model, however, does not capture the missing piece syndrome. 
de Souza e Silva \textit{et al}. \cite{scalabilityIssues} propose a protocol in which all the peers lacking only one piece, not necessarily the rarest piece, reduce their upload rate. Although this change in their closed system analysis where a departure causes an arrival leads to substantial gain in the maximum achievable throughput, the scheme requires fine-grained tuning.

Other related work is presented in \ref{subsection}. A detailed system model, the stability theorem for two pieces, the stability conjecture for arbitrary number of pieces, and a discussion of the proof technique are given in Section \ref{section:theoretical}. Decentralized variant of the group suppression protocol is defined in Section \ref{sect:decent}. Section \ref{sect:figures} offers simulation results confirming the stability theorem, the stability conjecture and the stability of the decentralized group suppression protocol. Section \ref{sect:bittorrent} contains the description and the simulations of BitTorrent modified to include the group suppression protocol. Finally, a discussion of contributions and findings as well as some future research directions is provided in Section \ref{sect:conclusion}.
\subsection{Other Related Work}\label{subsection}

Massouli{\'e} and Vojnovic \cite{Coupon} proved global stability in a model with non-persistent peers if all new peers arrive into the system with one uniformly random piece of the file uploaded by the fixed seed. The fixed seed's upload rate, however, may not be large enough to allow this in practice.

One of the earliest p2p models suitable to analytic analysis appeared in \cite{Qiu1}. The file consists of only one piece, which leads to a system that distinguishes peers based on whether they are seeds or not, rather than based on what fraction of the file they have. Furthermore, all incomplete peers are assigned the same effectiveness parameter and the incomplete peers exit the network according to a certain rate after becoming a seed, which means some of the peers are persistent. Under this model, they proved local stability of the associated deterministic system. Qiu and Sang \cite{Qiu2} later established global stability under the same model. The model in \cite{Qiu1} was extended by \cite{Tian} and \cite{BinFan}.

Considerable attention has been paid to improving the stability properties of p2p networks by connecting different \emph{swarms} where swarm refers to the set of all peers that want to download the same particular file \cite{StableSwarms, Zhou,  ContentBundling}. Bundling a number of files has been shown to increase the content availability and this benefit overcomes the burden of downloading unnecessary content for peers interested in unpopular files, which can be seen in reduced download time in \cite{ContentBundling}. Lacking any seeds, a network with non-persistent peers who aim to obtain some distinct files of size one was considered by Zhou \textit{et al}. in \cite{Zhou}. Peers with caches large enough to hold all the files enter the network with a file that some of the other peers are interested in. One notable result of the paper is that at most one swarm out of all swarms in the network is unstable. Zhu \textit{et al}.~\cite{StableSwarms} introduce a model  in which, unlike \cite{Zhou}, there exists a fixed seed, and non-persistent peers join the network empty, and the sets of pieces corresponding to distinct swarms may or may not be disjoint~\cite{StableSwarms}. In case of disjoint swarms, they determined that if the maximum of the arrival rates of all swarms are less than the upload rate of seed, then the overall system is stable and if it exceeds the upload rate of seed then the overall system is unstable. Zhang \textit{et al}. \cite{ZhangCoalitions} propose forming coalitions between peers that paves the way for different peer and piece selection policies that may surpass the performance of BitTorrent's peer and piece selection policies. 

\section{Problem Setup, The Group Suppression Protocol and Main Theorem}\label{section:theoretical}

We first describe the protocol used in Hajek and Zhu \cite{HajekMissing}.
We then define the group suppression protocol by describing how it
deviates from the Hajek and Zhu protocol.

In Hajek and Zhu's model \cite{HajekMissing}, which will be referred to as \textit{the unstructured p2p protocol}, the file to be downloaded is divided into $k$ pieces. Arrivals follow a Poisson process of rate $\lambda$. The new peers do not carry any pieces. The fixed seed remains in the system. All of the other peers are assumed to be non-persistent. The peer that initiates the contact uploads a piece to the other peer. In other words, the model employs a push policy instead of a pull policy. The uploading peer applies the "random useful piece selection" policy to determine the piece to upload, which amounts to selecting a piece among all useful pieces uniformly at random. If a contact occurs and the uploading peer has at least one piece that the other peer does not possess, we assume the upload of the randomly chosen piece takes place instantaneously. The time at which an incomplete peer contacts another peer is determined by Poisson processes with rate $\mu$ independent from an incomplete peer to incomplete peer. On the other hand, the seed contacts other peers according to a Poisson process of rate $U_s$. When it is time for a peer to contact another peer, it uniformly picks a peer from the set of all peers, which is called random peer selection policy.

We now build the group suppression protocol upon the unstructured p2p protocol by employing two modifications to it. The first and arguably less significant modification requires the fixed seed to select a peer uniformly at random among the peers with the least number of pieces rather than uniformly at random among all peers and to upload a random useful piece. This seed policy is usually called the most deprived policy. It is sometimes combined with another rule that the seed upload the rarest piece rather than a random piece, which implies the seed has access to every peer's piece profile in the network \cite{scalabilityIssues, deSouza}. Our group suppression protocol allows the seed to upload a random useful piece.

To understand the second modification, recall from Section \ref{sect:introduction} that two peers belong to the same group if their piece profiles exactly match each other and the strictly most populous group is designated as the largest club. We assume for now that peers have access to this central knowledge. Notice that the network may not necessarily have a largest club. The backbone of the group suppression protocol relies on preventing a group of peers from dominating the network. In order to accomplish that, the second deviation from the unstructured p2p protocol, which earns the group suppression protocol its name, requires the peers in the largest 
club to deny the upload to any peer with a smaller or equal number of pieces whenever they initiate a contact with such a peer. 

The group suppression protocol contributes to stability by curbing the growth of the largest club and by concentrating the seed's upload on the peers with the least number of pieces. Since the largest club rapidly recruits other peers once it starts to dominate the network, stopping the largest club requires eradication of its source of growth, which is uploads from the largest club to peers who hold a subset of the largest club's pieces, or more generally, 
all the peers except the ones holding a greater portion of the file. 
The peers with greater number of pieces than the largest club peers cannot be absorbed into the largest club. Therefore, the largest club is allowed to utilize its upload capacity for these peers but not for others. The protocol's rule concerning the seed originates from the intuition that if the seed uploads a rare piece to a peer, it is preferable that the peer be one who will remain in the system for as long as possible. A peer with the least number of pieces is presumably more likely to stay longer in the network than other peers.

We state the theorem regarding the stability of the group suppression protocol for $k=2$ and also conjecture the stability of the protocol for arbitrary $k$.
\\\\\\
\textbf{Main Theorem:} If the group suppression protocol is employed and $k=2$, then the continuous-time Markov chain representing the network is positive recurrent for any $(\lambda > 0, U_s > 0, \mu > 0)$. That is, the p2p network with the group suppression protocol and $k=2$ is stable for any $(\lambda > 0, U_s > 0, \mu > 0)$.
 \\\\
\textbf{Conjecture 1:} Under the group suppression protocol for arbitrary $k$, the continuous-time Markov chain representing the network is positive recurrent for any $(\lambda > 0, U_s > 0, \mu > 0)$. That is, the p2p network with the group suppression protocol and arbitrary $k$ is stable for any $(\lambda > 0, U_s > 0, \mu > 0)$.
 \\\\
\textbf{Sketch of the proof of Main Theorem:}\\
We employ the Foster-Lyapunov proposition to prove positive recurrence of the underlying Markov chain. To state the proposition, we need to define the drift of a potential function $V(\boldsymbol{s})$:
\begin{equation}
DV(\boldsymbol{s}) = \sum_{\boldsymbol{y}:\boldsymbol{y}\neq \boldsymbol{s}} q(\boldsymbol{s},\boldsymbol{y})[V(\boldsymbol{y})-V(\boldsymbol{s})]\label{drift},
\end{equation}where $q(\boldsymbol{s},\boldsymbol{y})$ is the transition rate from state $\boldsymbol{s}$ to $\boldsymbol{y}$.\\\\
\textbf{Foster-Lyapunov Proposition}~\cite{HajekBook}\textbf{:} Let $\boldsymbol{\phi}$ be a time homogeneous, irreducible and continuous time Markov process and $\boldsymbol{S}$ be its state space. If there exists a finite set of states $\boldsymbol{C} \subset \boldsymbol{S} $, a potential function $V(\boldsymbol{s}):\boldsymbol{S}\rightarrow(0,\infty)$ and some constants $b>0$, $\epsilon>0$, such that:
\begin{eqnarray}
\{ \boldsymbol{s}:V(\boldsymbol{s}) < K \} ~~\text{is finite } \forall K \label{foster:fst2} \\
DV(\boldsymbol{s}) \leq -\epsilon + b\boldsymbol{1}_C (\boldsymbol{s})~~\forall \boldsymbol{s} \in \boldsymbol{S} \label{foster:fst3}
\end{eqnarray}
then $\boldsymbol{\phi}$ is positive recurrent.\\\\
First note that the underlying Markov process for p2p network at hand is time homogeneous, irreducible and continuous time. To describe the particular potential function that satisfies Foster-Lyapunov proposition, we introduce the following notation for the system with $k = 2$ pieces: \\\\
\emph{type 0} peer: a peer that has no pieces \\
\emph{type~1} peer: a peer that has piece 1 but not piece 2\\
\emph{type 2} peer: a peer that has piece 2 but not piece 1\\
$n_0$ : the number of \emph{type 0} peers.\\
$n_1$ : the number of \emph{type 1} peers.\\
$n_2$ : the number of \emph{type 2} peers. \\
$\boldsymbol{s}$: the state $(n_0, n_1, n_2)$.\\
$s$ : the number of peers in the system.\\
It is convenient to consider the fixed seed as not part of the system. Therefore, we have $s=n_0+n_1+n_2$\\
We choose the potential function as follows:
\begin{eqnarray} 
V=(a^+)^2 + (b^+)^2 + d^2  \label{potential:eqn1},
\end{eqnarray}where 
\begin{align*}
&a=n_0+\min(n_0,n_1)+c_1(n_1-n_0)^+ - c_2n_2 \\
&b=n_0+\min(n_0,n_2)+c_1(n_2-n_0)^+ - c_2n_1 \\
&d=c_3n_0+c_4n_1+c_4n_2.
\end{align*}

In the Appendix, where we postpone the rest of the proof of the Main Theorem, we specify a finite set of states $\boldsymbol{C}$ and a set of conditions on constants $c_1, c_2, c_3, c_4, p$ for every $(\lambda, U_s, \mu)$, and then show that under the group suppression protocol, the potential function $V(\boldsymbol{s})$ in \eqref{potential:eqn1} and the specified $\boldsymbol{C}$ satisfies Foster-Lyapunov Proposition for any $(\lambda>0, U_s>0, \mu>0)$.

We now discuss the choice of the potential function $V(\boldsymbol{s})$ in \eqref{potential:eqn1}.  As the complete reasoning behind the construction of the potential function is rather involved, consider a simple scenario where $c_1=1, c_2=1, c_3=1, c_4=1$, although this choice of constants violates the conditions imposed on these constants in the Appendix. Then, the elements of the potential function become:
\begin{align*}
a=n_0+n_1-n_2\\
b=n_0+n_2-n_1\\
d=n_0+n_1+n_2.
\end{align*}
Then $a$ is the \emph{excess demand} for piece 2, i.e., the number of 
peers that demand piece 2 minus the number of peers that can supply it.
Likewise, $b$ is the excess demand for piece 1. The $a$ and $b$ terms
in the Lyapunov function therefore penalize lopsided network
states in which there is a dominant largest club. The $d$ term
evidently penalizes the overall network size.
\section{The Decentralized Group Suppression Protocol}
\label{sect:decent}
The group suppression protocol might appear to require a certain level of centralization because the seed must find the peers with smallest number of pieces in the system, and every peer needs to determine if they belong to the largest club when they contact another peer. However, both elements of centralization in the protocol can be relaxed for practical implementations of p2p systems. To remove centralization, we now define \textit{the decentralized group suppression protocol}. The protocol is built on the unstructured p2p protocol as follows:  Every incomplete peer considers the piece profiles of the last three peers it has contacted along with its own piece profile. If the multiplicity of its own piece profile is strictly more numerous than that of any other piece profiles in this set, then it designates itself as a largest club member.
Note that any peer will always have at least two piece profiles including its own piece profile in that set. 
Second, the fixed seed receives the id of peers upon their arrival and stores the most recent $5$ arrivals' ids, the most recent ranking first, and uploads to the highest ranking peer that is still in the network. If all $5$ peers appear to have left and no arrival has happened yet, the fixed seed picks a peer uniformly at random among all peers. Thus, the decentralized group suppression protocol handles the largest club membership decision in a distributed fashion for incomplete peers and offers the fixed seed a decentralized way of finding the peers that are more likely to be among the peers with the smallest portion of the file. It should be noted that the effectiveness of both features of the protocol rely on the uniform peer selection policy.

\section{Simulations for the Main Theorem and Conjecture 1}
\label{sect:figures}
\begin{figure*}
\centering
\begin{minipage}{.47\textwidth}
  \centering
  \includegraphics[width=0.9\linewidth]{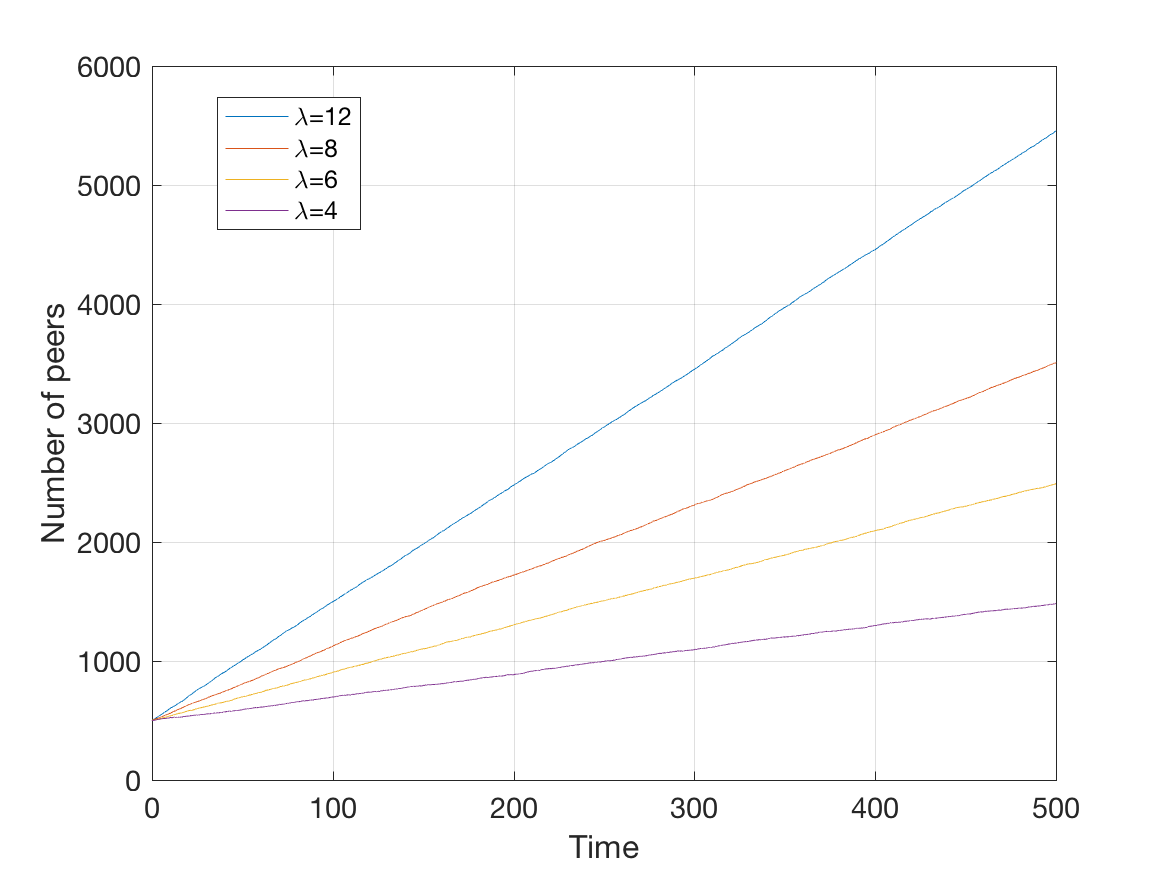}
  \captionof{figure}{The total number of peers in the network with the unstructured p2p protocol when $k=2$ and $\lambda$ is varying.}
  \label{fig:test1}
\end{minipage}\hfill
\begin{minipage}{.47\textwidth}
  \centering
  \includegraphics[width=0.9\linewidth]{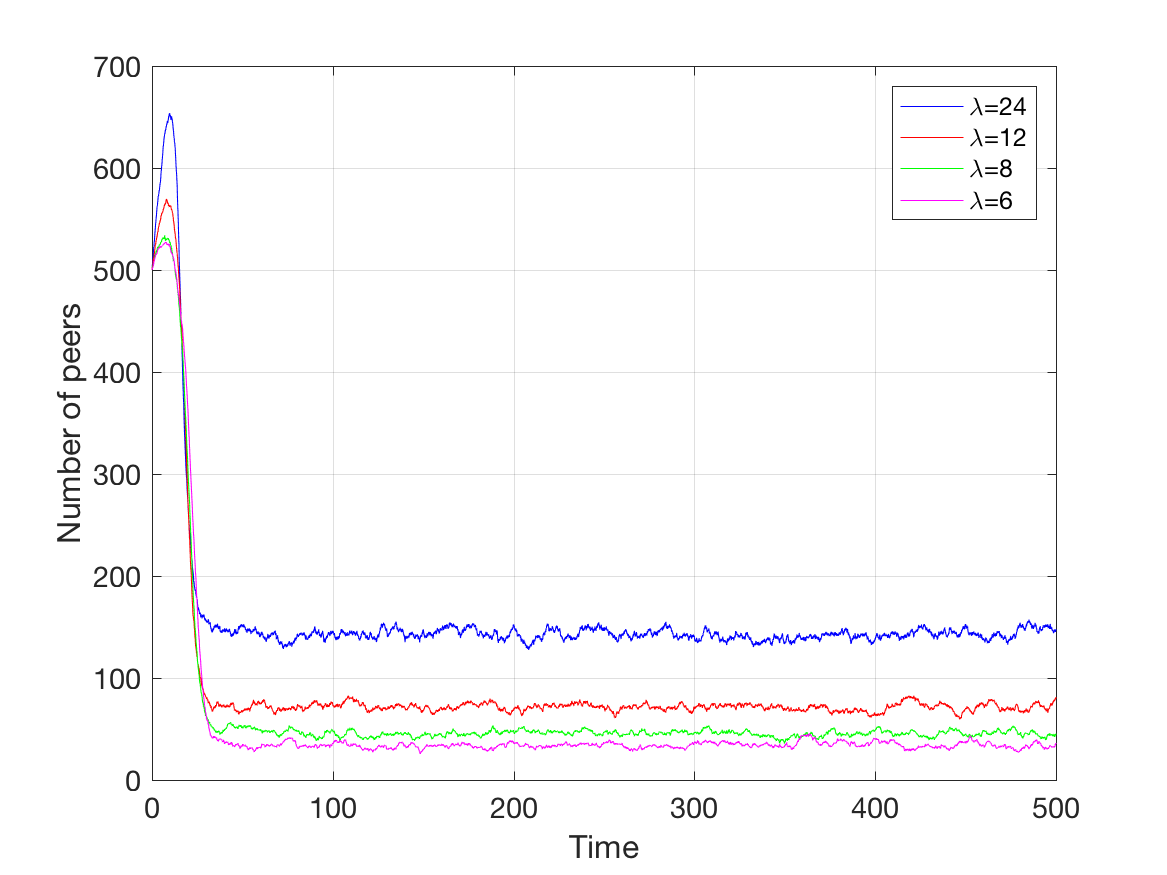}
  \captionof{figure}{The total number of peers in the network with the group suppression protocol when $k=2$ and $\lambda$ is varying.}
  \label{fig:test2}
\end{minipage}%
\end{figure*}

\begin{figure*}
\centering
\begin{minipage}{.47\textwidth}
  \centering
  \includegraphics[width=.9\linewidth]{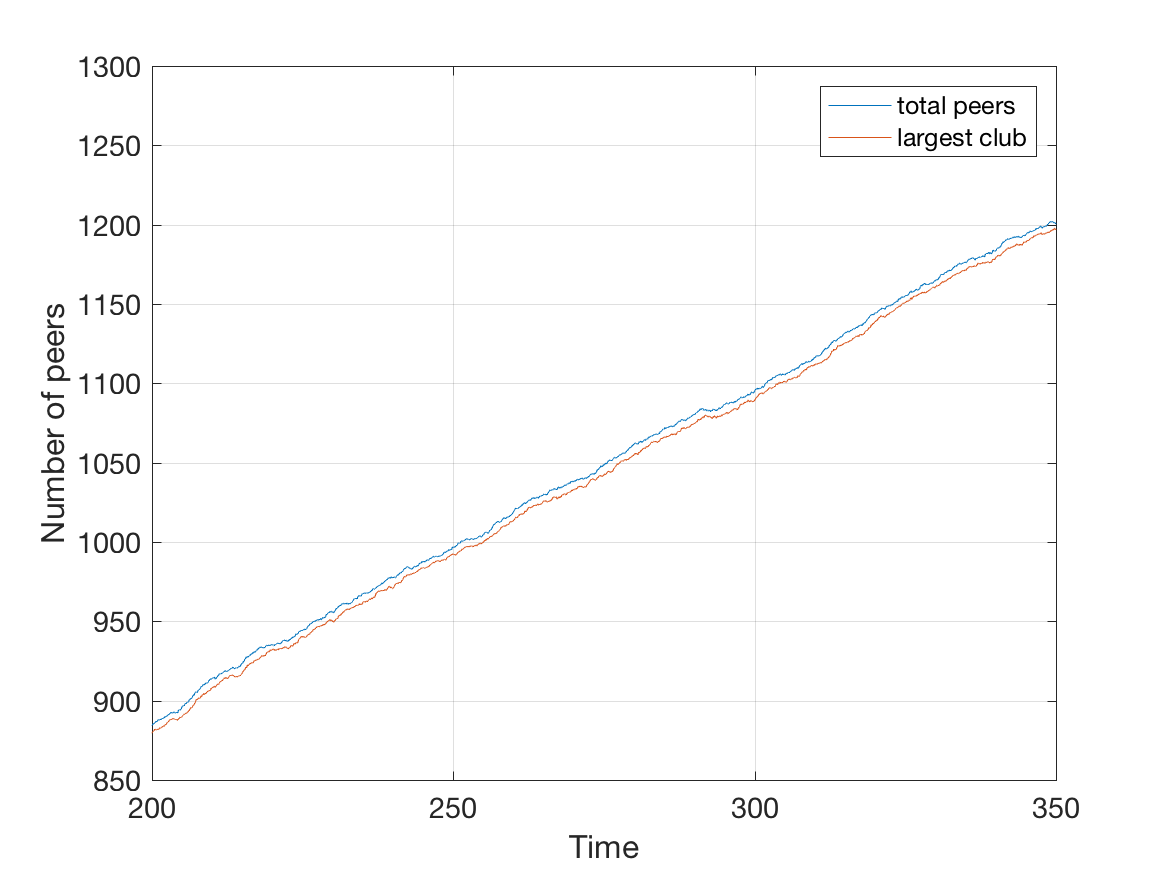}
  \captionof{figure}{Comparison of the total number of peers to population of the largest club peers in the network with the unstructured p2p protocol when $k=2$ and $\lambda=4$.}
  \label{fig:test3}
\end{minipage}\hfill
\begin{minipage}{.47\textwidth}
  \centering
  \includegraphics[width=.9\linewidth]{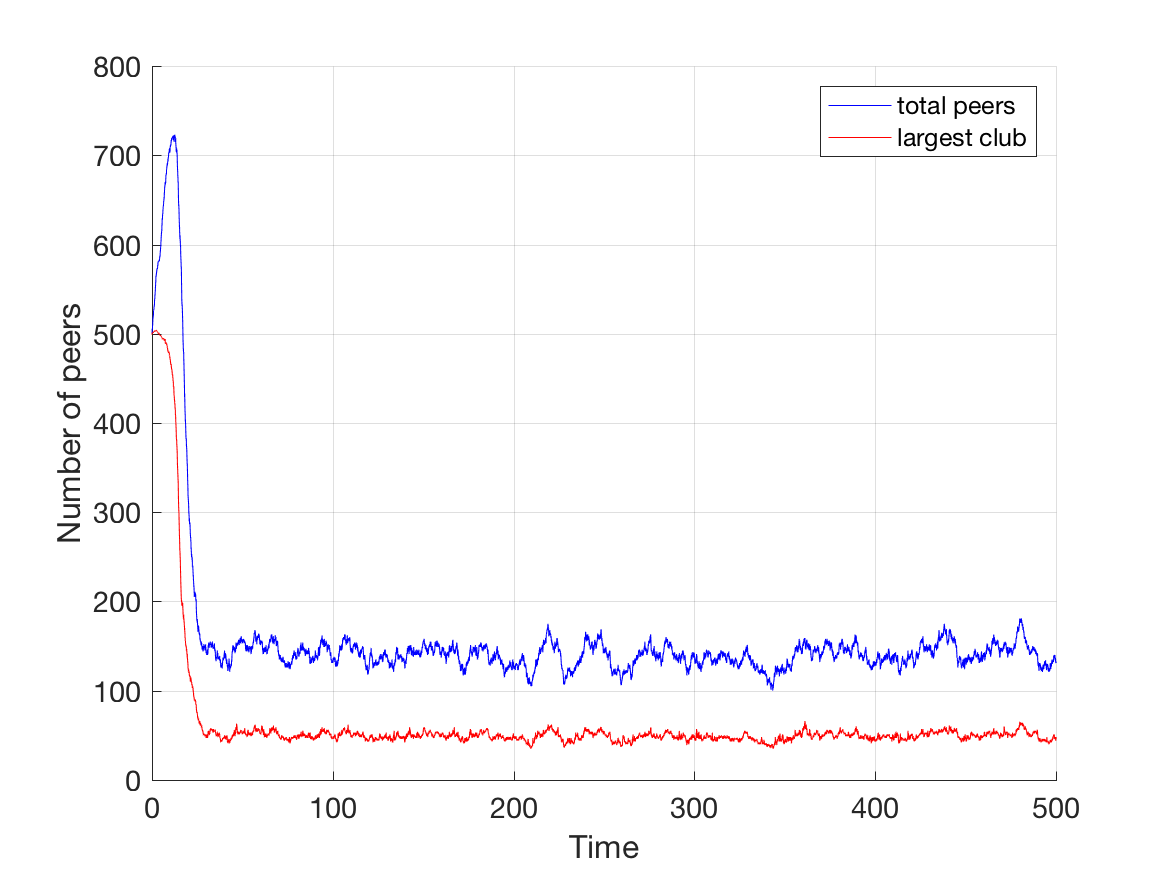}
  \captionof{figure}{Comparison of the total number of peers to population of the largest club peers in the network with the group suppression protocol when $k=2$ and $\lambda=24$.}
  \label{fig:test4}
\end{minipage}%
\end{figure*}

This section simulates the unstructured p2p protocol and the group suppression protocol defined in Section \ref{section:theoretical}, and the decentralized group suppression protocol defined in Section \ref{sect:decent}. 

The following initial conditions and specifications of rates apply to both protocols: The seed contacts the other peers according to a Poisson process of rate 2. Every  incomplete peer's contact process follows according to an independent Poisson process of rate 1. The seed's upload rate is chosen to be 2 instead of 1 because such choice captures the fact that the seed does not have to download anything and, therefore, it is likely to allocate more bandwidth to the upload than the incomplete peers. Moreover, increasing the seed's upload rate makes it more challenging for the largest club to pull the network towards instability, which nevertheless will happen when the unstructured p2p protocol is simulated. The simulations start with $500$ peers at $t=0$ and all of the peers except the fixed seed possess all the pieces of the file but the one common piece. And all incomplete peers depart the network as soon as they obtain the entire file. All simulations were done in Matlab.

\begin{figure*}
\centering
\begin{minipage}{.47\textwidth}
  \centering
  \includegraphics[width=.9\linewidth]{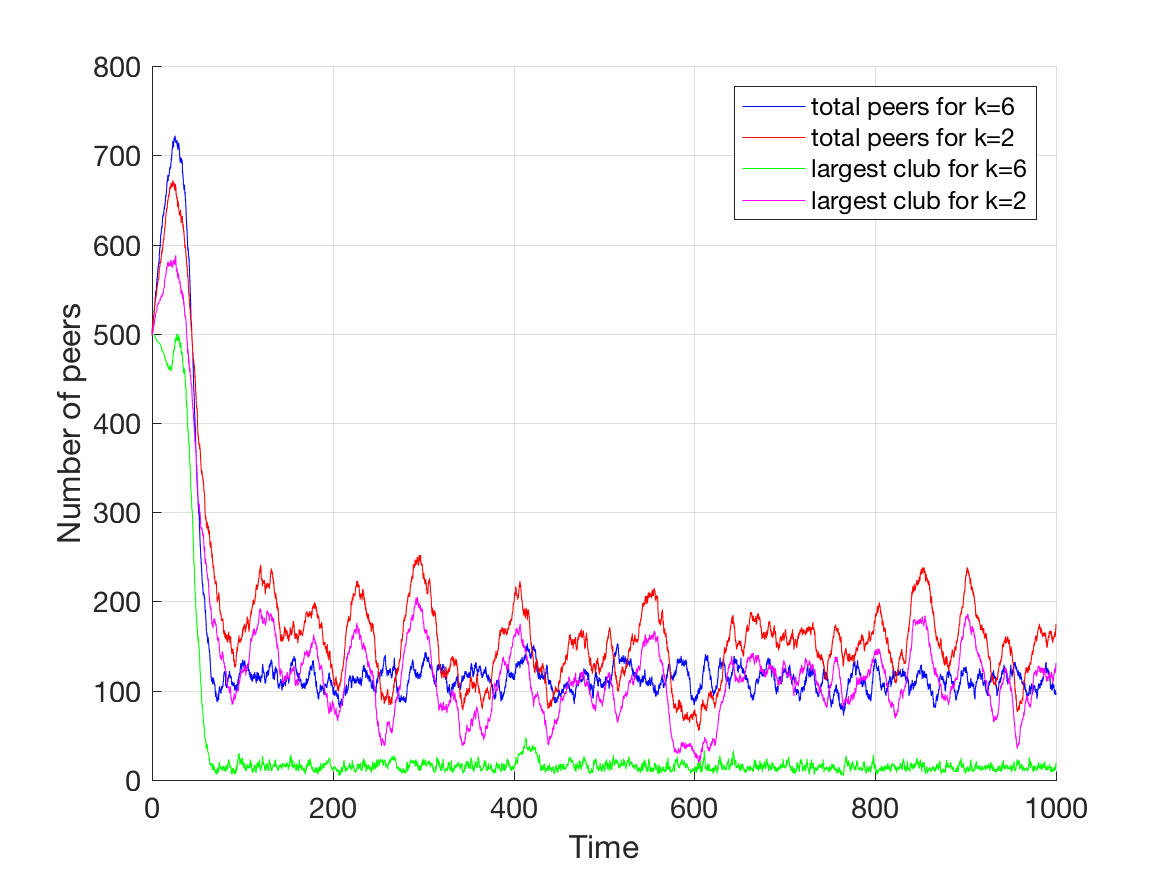}
  \captionof{figure}{Comparison of the total number of peers to the largest club when the decentralized group suppression protocol is in effect with $\lambda=12$ and $k=2, 6$}
  \label{fig:test42}
\end{minipage}\hfill
\begin{minipage}{.47\textwidth}
  \centering
  \includegraphics[width=.9\linewidth]{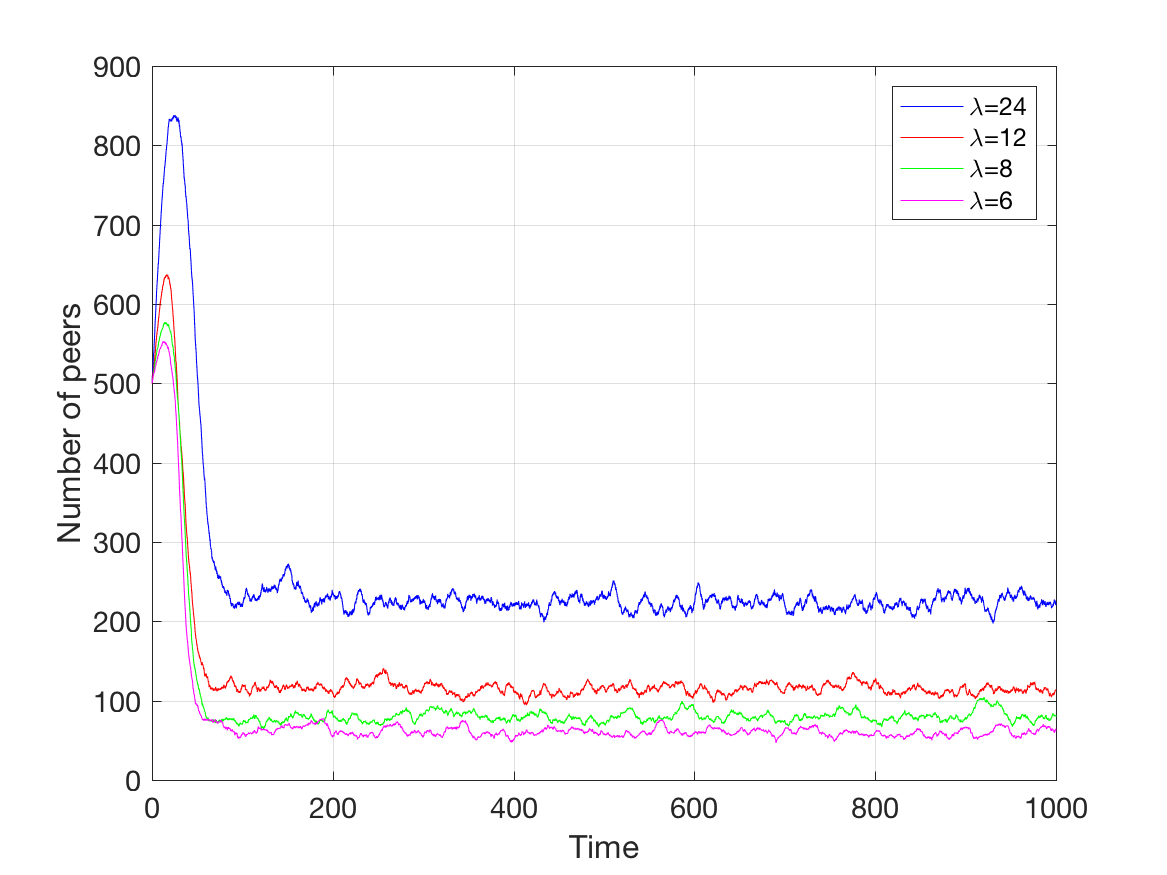}
  \captionof{figure}{Population of the p2p network with the decentralized group suppression protocol when $k=6$ and $\lambda$ is varied}
  \label{fig:test43}
\end{minipage}%
\end{figure*}

Fig. \ref{fig:test1}-\ref{fig:test4} are intended to verify the Main Theorem and therefore the file is divided into only two pieces. Instability is seen in Figure \ref{fig:test1} where the unstructured p2p protocol is applied and $\lambda$ is varied. Since the arrival rate $\lambda$ is greater than the seed's upload rate $U_s$, instability is expected and the result is consistent with the main theorem of Hajek and Zhu \cite{HajekMissing}. In Figure \ref{fig:test1}, the networks roughly grow with rate $\lambda-U_s$, which is consistent with the understanding that almost no peer except the seed can help another exit the system by supplying the missing piece. In fact, Figure \ref{fig:test3} bears more direct evidence for the claim as the line that represent the number of the largest club peers closely trails the line that represents the total number of peers in the system. The total number of peers is shown in Figure \ref{fig:test2} for varying $\lambda$ values when the group suppression protocol is applied. The networks for varying arrival rates, which are larger than the upload rate of the seed, recover from very skewed piece distribution and achieve stability. Figure \ref{fig:test4} shows both the total number of peers and the largest club population when $\lambda=24$. The apparent increases in the population of the largest club apparently give rise to the sudden jumps in the total number of peers. In the same vein, the downward trend in the population of the largest club is followed by the same trend in the total number of peers. Between the growth and the decline of the largest club, the group suppression protocol cuts off the supply of pieces from the largest club to any peer who does not hold a greater portion of the file, giving those peers more time to obtain and distribute the rarest piece without being absorbed into the largest club. Note that while Fig. \ref{fig:test4} and \ref{fig:test42} demostrate single realizations,
all of the other figures in this section represent an average of five individual realizations of their corresponding stochastic network. 

\begin{figure*}
\centering
\begin{minipage}{.47\textwidth}
  \centering
  \includegraphics[width=.9\linewidth]{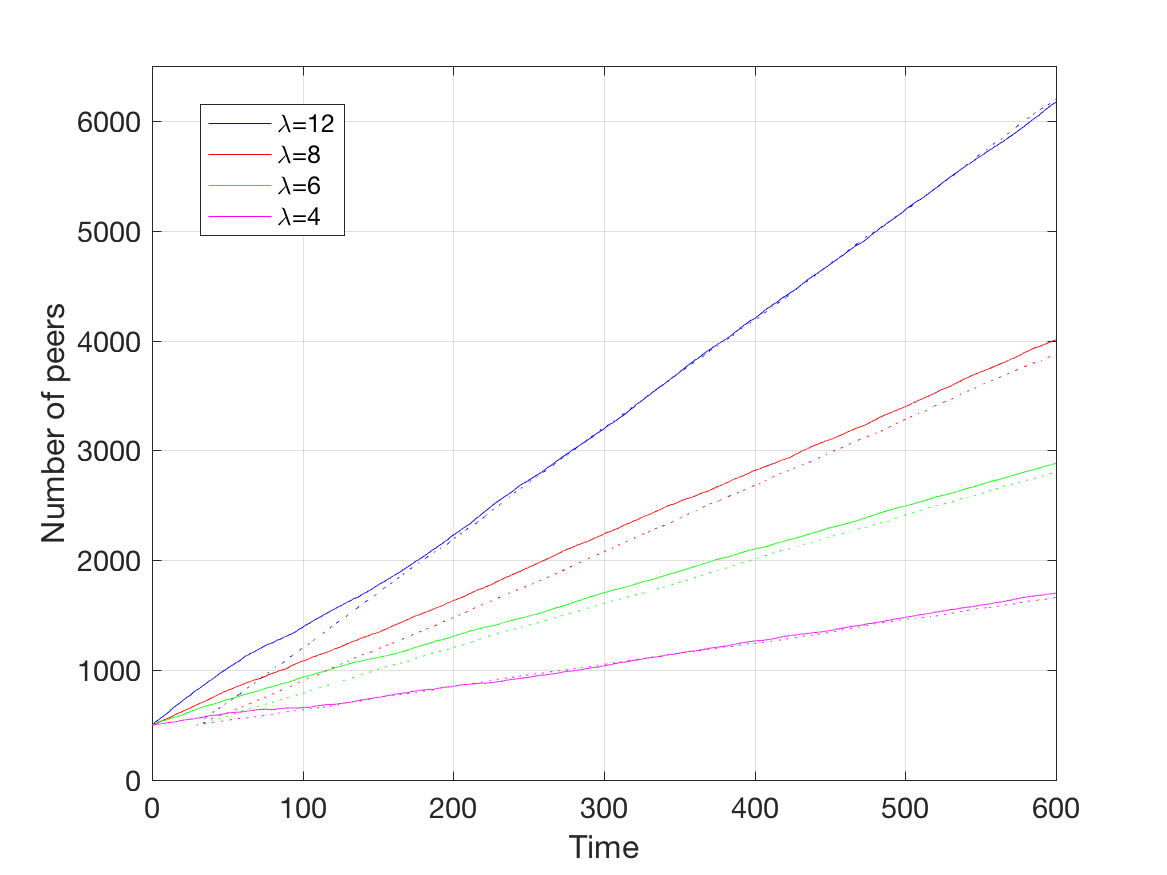}
  \captionof{figure}{When the unstructured p2p protocol is in effect with varying $\lambda$, the total number of peers are represented in solid line for $k=48$ whereas the dashed line is a 30-time unit delayed version of the total number of peers for $k=6$.}
  \label{fig:test7}
\end{minipage}\hfill
\begin{minipage}{.47\textwidth}
  \centering
  \includegraphics[width=.9\linewidth]{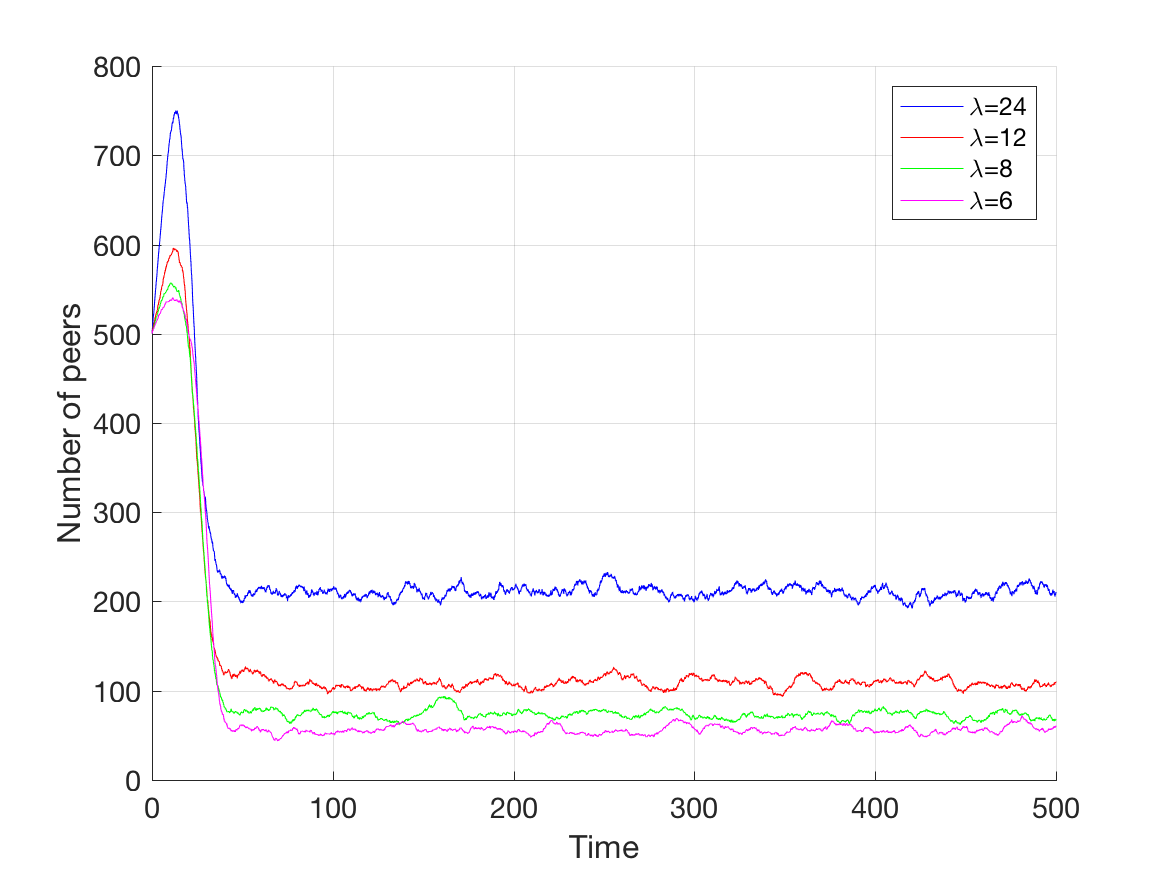}
  \captionof{figure}{The total number of peers in the p2p network with the group suppression protocol when $k=6$ and $\lambda$ is varied.}
  \label{fig:test9}
\end{minipage}%
\end{figure*}
\begin{figure*}
\centering
\begin{minipage}{.47\textwidth}
  \centering
  \includegraphics[width=.9\linewidth]{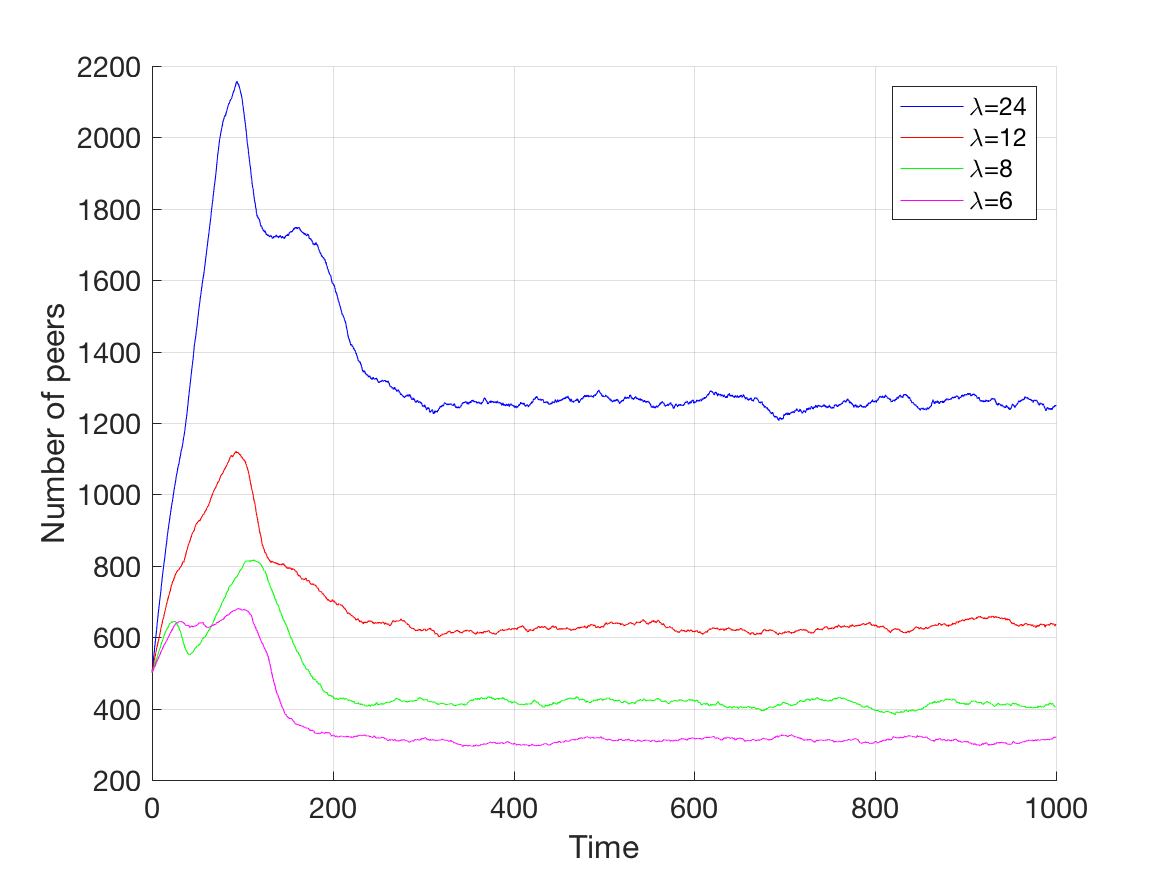}
  \captionof{figure}{Population of the p2p network with the group suppression protocol when $k=48$ and $\lambda$ is varied.}
  \label{fig:test10}
\end{minipage}\hfill
\begin{minipage}{.47\textwidth}
  \centering
  \includegraphics[width=.9\linewidth]{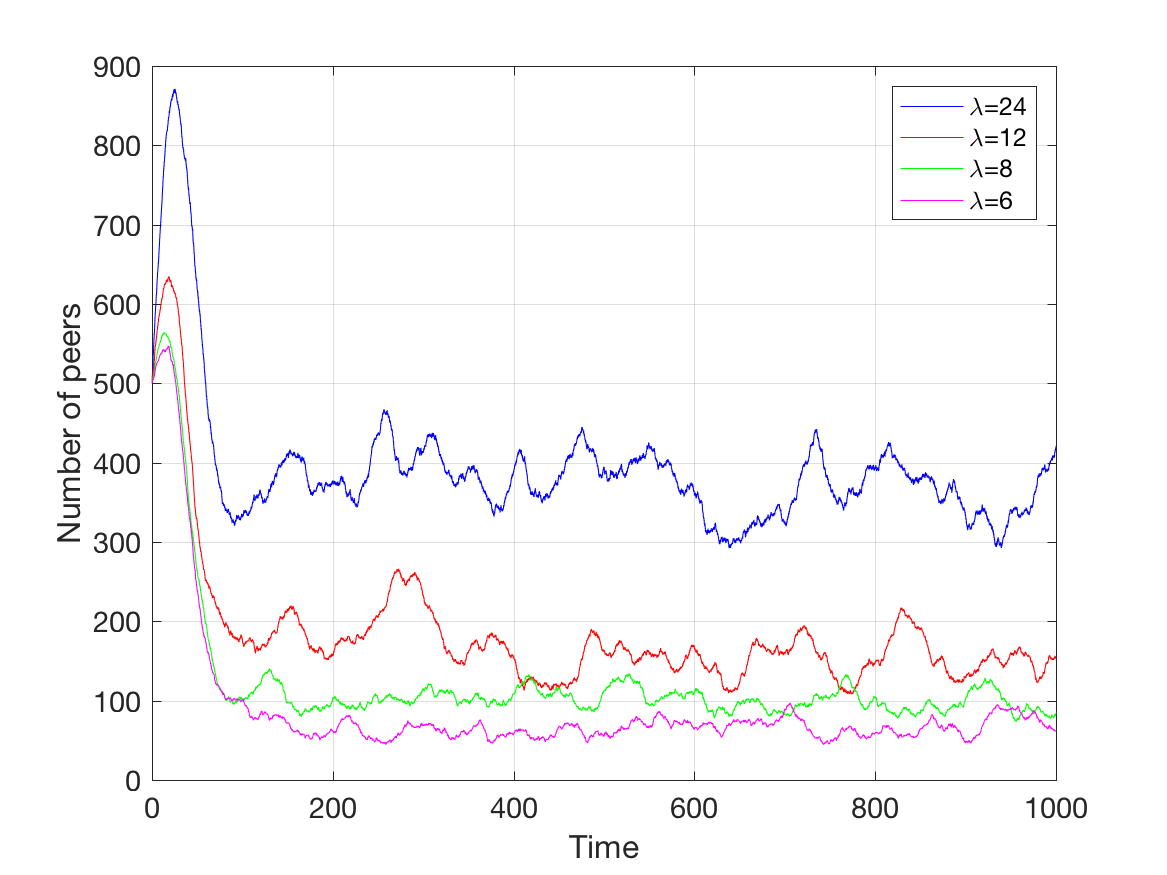}
  \captionof{figure}{Population of the p2p network with the decentralized group suppression protocol when $k=2$ and $\lambda$ is varied}
  \label{fig:test41}
\end{minipage}%
\end{figure*}

\begin{figure*}
\centering
\begin{minipage}{.47\textwidth}
  \centering
  \includegraphics[width=.9\linewidth]{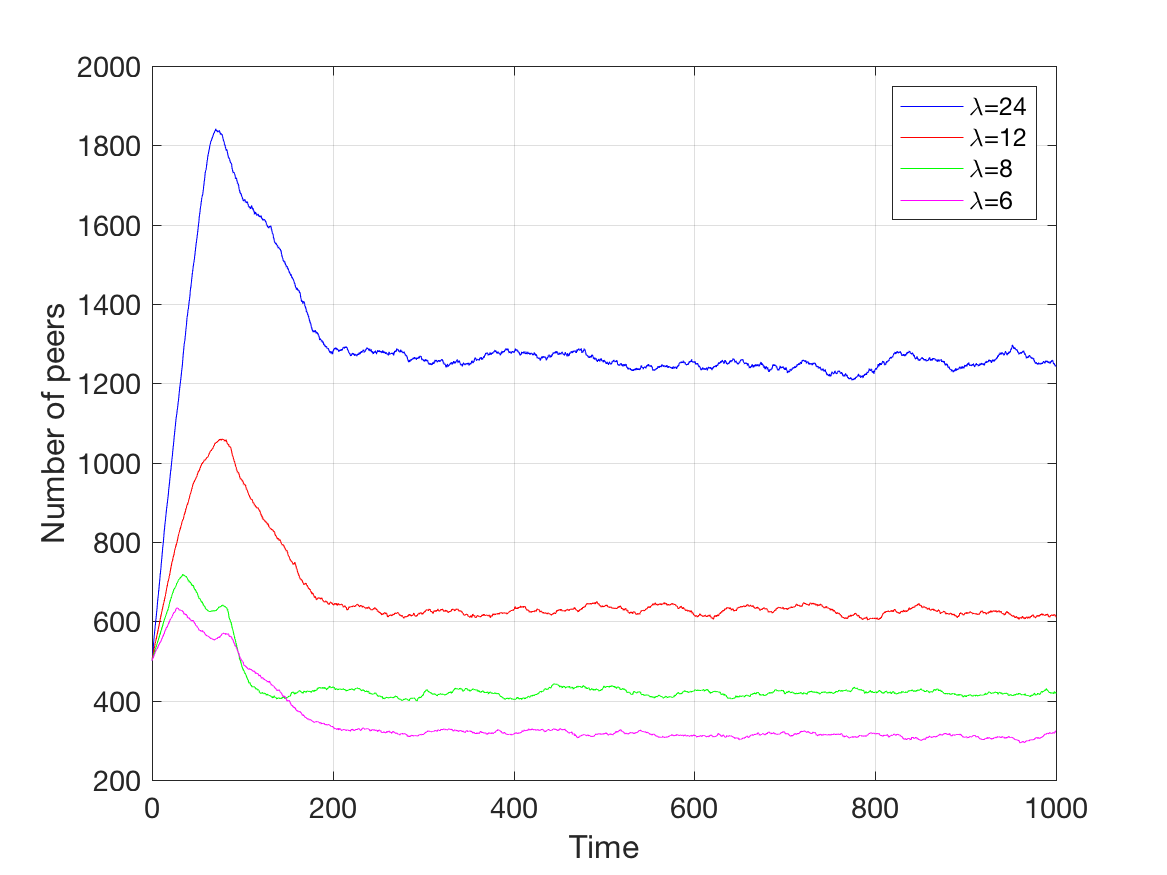}
  \captionof{figure}{Population of the p2p network with the decentralized group suppression protocol when $k=48$ and $\lambda$ is varied}
  \label{fig:test44}
\end{minipage}\hfill
\begin{minipage}{.47\textwidth}
  \centering
  \includegraphics[width=.9\linewidth]{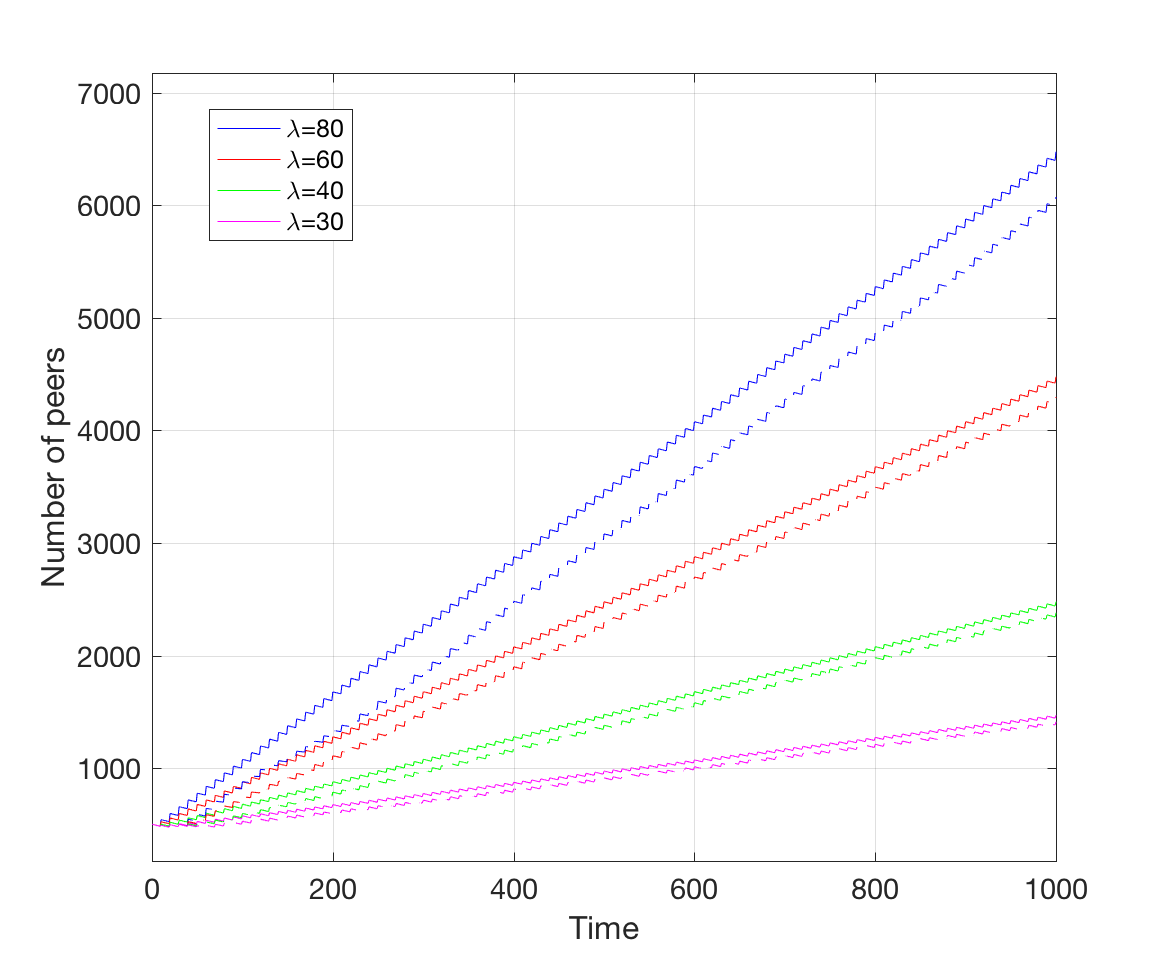}
  \captionof{figure}{In the network governed by the BitTorrent-like protocol, the total number of peers are represented in solid lines for $k=12$ while the dashed lines show 30-second delayed versions of the total number of peers for $k=48$.}
  \label{fig:test13}
\end{minipage}%

\end{figure*}

\begin{table*}
\centering 
%\resizebox{\columnwidth}{!}{ 
%\resizebox{1.0\linewidth}{!}{ 
%{adjustbox}{width=\columnwidth}{4}{
%\scalebox{1.2}{
%\begin{tabular}{| l | c | c | c | c |}
\begin{tabular}{| K{4.7cm} | K{2.8cm} | K{2.8cm} | K{2.8cm} | K{2.8cm}|}
\hline
 Protocol & Average sojourn time for $k=25$  & Average sojourn time for $k=50$  & Average sojourn time for $k=100$  \\ [0.5ex]
\hline 
\text{Group suppression protocol} & 28.95 &  54.50  & 106.11 \\
\hline
\text{Decentralized group suppression protocol} & 29.12 & 54.60 & 105.39  \\
\hline
\text{Waiting protocol \cite{HajekWaiting}} & 29.18 & 54.92 & 105.69  \\
\hline
\text{Forced Friedman protocol \cite{Reittu}} & 30.54 & 55.89 & 105.84   \\
\hline
\text{Common chunk protocol \cite{Venkat}} for m=5 & 30.39 & 57.63 &  110.92      \\
\hline 
\text{Common chunk protocol \cite{Venkat}} for m=3 & 36.88 & 67.39 & 126.02  \\ [1ex] 
\hline
\end{tabular}
%}
\caption{Average sojourn times for different protocols when $\lambda = 6$ and $k$ is varied.}
\label{table:sojourn} % is used to refer this table in the text
\end{table*}
Conjecture 1 in Section \ref{section:theoretical} states that any p2p network with the group suppression protocol is stable regardless of the number of the pieces the file is divided into. Figure \ref{fig:test7}, where the unstructured p2p protocol is in effect, provides experimental evidence that when $\lambda$ is greater than $U_s$, the missing piece syndrome phenomenon develops irrespective of the number of pieces $k$. Furthermore, the number of pieces in the file appears not to affect the slope of the total number of peers in the network because the total number of peers roughly grows with $\lambda-U_s$ per time unit for any $k$ in Figure \ref{fig:test7}. The group suppression protocol is put into practice in Fig. \ref{fig:test9}-\ref{fig:test10} for various ($\lambda, k$) pairs. The stability follows a transient period and prevails permanently in all of different  $(\lambda,k)$ pairs. Therefore, Fig. \ref{fig:test9}-\ref{fig:test10} provide evidence that the stability conjecture holds true. 

The simulations shown in Fig. \ref{fig:test41}-\ref{fig:test44} indicate that the decentralized group suppression protocol is stable for all $(\lambda,k)$ pairs considered. The plots regarding the decentralized group suppression protocol are mostly very similar to their group suppression counterparts. The only outliers are Fig. \ref{fig:test41}-\ref{fig:test42}. First of all, the decentralized group suppression protocol in Fig. \ref{fig:test41}, where $k=2$ and $\lambda$ is large, appears to be more prone to spikes compared to the same setting with the group suppression protocol in Fig. \ref{fig:test2} even though we take the average of five realizations. It stems from the higher probability of error in deciding the largest club membership due to the small number of type of peers in the network. Comparing Fig. \ref{fig:test4} to Fig. \ref{fig:test42}, both of which demonstrate single realizations, underlines the difficulty of controlling the largest club when $k=2$, which signals the largest club decision mistakes affect the network so as to create substantial ripples. Nevertheless, the largest club of the decentralized group suppression protocol behaves smoothly in Fig. \ref{fig:test42} when $k$ is increased to 6. Since $k$ is usually large and the decentralized group suppression's goal is intended to be deployed for practical use, the spikes observed for $k=2$ should not constitute a basis for concern. Indeed, none of the figures of the decentralized group suppression protocol for $k$ larger than 2 has no such noticeable spikes in Fig. \ref{fig:test43}-\ref{fig:test44}.

After stability, the arguably most important measure of p2p system
performance is the mean sojourn time. We compare different protocol's sojourn time performances to that of the group suppression protocol and to that of the decentralized group suppression protocol in Table 1. One of the other protocols is due to Zhu and Hajek~\cite{HajekWaiting}, which is referred to as the waiting protocol in Table 1. This protocol assumes the peers stay in the network on average long enough to upload one piece after they become seeds in our implementation of the waiting protocol. The second protocol in the comparison is O\v{g}uz \textit{et al}.'s common chunk protocol in \cite{Venkat}, which is simulated for both when the number of the contacts $m$ that peers lacking only one piece make is three and five. The final protocol simulated is Reittu's forced Friedman protocol \cite{Reittu}, which requires every peer to sample three other peers and allows the peer to download a random piece from the set of pieces that show up exactly once in the aggregate piece profile of these three sampled peers. If the set turns out to be empty, the peer skips without downloading any piece. We obtained $100$ individual realizations for each protocol where the simulations started with $499$ peers holding all pieces but one common piece and a fixed seed, and $k=25, 50, 100$ and $\lambda=6$. We also set $U_s=\mu=1$  because the common chunk protocol and forced Friedman protocol do not distinguish between the upload capacity of a seed and an incomplete peer. In every realization, we collected the sojourn times of the first $500$ peers to exit the system after we had waited $2000$ time units, which we observed was enough for all networks to reach steady-state. The results in Table 1 show that the first four protocols including the group suppression protocol and the decentralized group suppression protocol have very close average sojourn times for all $k$ values. The average sojourn time of the common chunk protocol for both $m=3,5$ do not scale as well with $k$ compared to the other protocols. This could give rise to considerable sojourn time performance loss when the file is divided into tens of thousands of pieces, which is a common regime in real-life p2p networks. On the other hand, although the average sojourn times of both the waiting protocol and the forced Friedman protocol are on par with that of the group suppression protocol and the decentralized group suppression protocol, it is important to recall that the former relies on persistent peers and the latter depends on non-opportunistic peers. Both the group suppression protocol and the decentralized group suppression protocol assume non-persistence and opportunistic behavior. It is also notable that decentralizing the group suppression protocol does not lead to a sojourn time performance penalty.

\section{Models of the modified BitTorrent protocols and their simulations}\label{sect:bittorrent}
The section is organized as follows: First, the official BitTorrent protocol's fundamental algorithms, namely the rarest first and the unchoking algorithms, are briefly explained. Then a simplified yet rather loyal implementation of the official BitTorrent protocol, which lends itself to simulations and we call the \textit{BitTorrent-like protocol}, is introduced. We then construct \textit{the BitTorrent-like protocol with group suppression} by incorporating the features of the group suppression protocol into the BitTorrent-like protocol. The section is concluded with simulation results of the BitTorrent-like protocol and the BitTorrent-like protocol with group suppression, all of which were done in Matlab.

Since BitTorrent's source code is no longer public, we consider one of the older versions of BitTorrent, namely version 4.20.0. In the official BitTorrent protocol, every peer maintains a set of peers, which are usually called the neighbor peers. The number of the neighbor peers cannot exceed some particular number. If a peer's number of neighbor peers falls below a pre-defined threshold, then that particular peer requests new peers from the tracker of the torrent. The file is divided into pieces and each piece is broken into a certain number of blocks. A peer cannot offer a partial piece. Furthermore, every peer knows the individual piece profiles of its neighbor peers at any time although they do not possess such knowledge for other peers in the network.  Peer A is said to be unchoked by Peer B if Peer B decides to fulfill any download request Peer A may make. Peer A is called an interested peer by Peer B if Peer B has some piece that Peer A may want to request. The rarest first policy is employed, which simply means that when the peer is unchoked by one of its neighbors, the peer first requests from the neighbor the rarest piece among the pieces needed by the peer. The rarest piece is determined by the downloading peer based on all of its neighbors' piece profiles. Both the seed and incomplete peers could upload to a certain number of peers, which is  at most and by default four. The unchoking policy varies based on whether the seed or an incomplete peer is unchoking. In every unchoking round of 10 seconds, an incomplete peer ranks all the interested neighbor peers in a descending order based on their upload rates to this incomplete peer and unchokes the first three peers. Once in every three rounds, an incomplete peer selects one additional peer among its neighbors uniformly at random and keeps it unchoked for three rounds, which is called optimistic unchoking. On the other hand, the seed ranks all the unchoked interested neighbor peers that have been unchoked in last 20 seconds  based on when they have been unchoked, the latest being the first. In the first two rounds of seed's unchoking, it allocates one of its four upload slots to a uniformly chosen peer and the remaining three slots are filled from the top of the list of lately unchoked peers. In the last of every three rounds, all four unchoked peers are kept. The rarest first and  the unchoking rules are the main policies of BitTorrent. Yet, there exists other rules governing BitTorrent networks such as strict priority, random first and endgame mode. Strict priority refers to requesting the remaining blocks of a partial block before making a request on the blocks of a new piece. According to random first policy, peers with less than four pieces select the piece to download randomly rather than based on rarity. Then they switch to the rarest first policy when they obtain four full pieces. Endgame mode kicks in when all the blocks have been requested by a peer. The mode enables the peer to request all the remaining blocks from every neighbor that is unchoking the peer and it also allows the peer to be interested in all the neighbors for the blocks that are currently choking it. Although the explanation of the official BitTorrent protocol here should be sufficient to understand the simulations, a reader is advised to consult \cite{Cohen}, \cite{LegoutBitTorrentDetails} and \cite{LegoutStability} for a deeper understanding of the official BitTorrent protocol.

\begin{figure*}
\centering
\begin{minipage}{.47\textwidth}
  \centering
  \includegraphics[width=.9\linewidth]{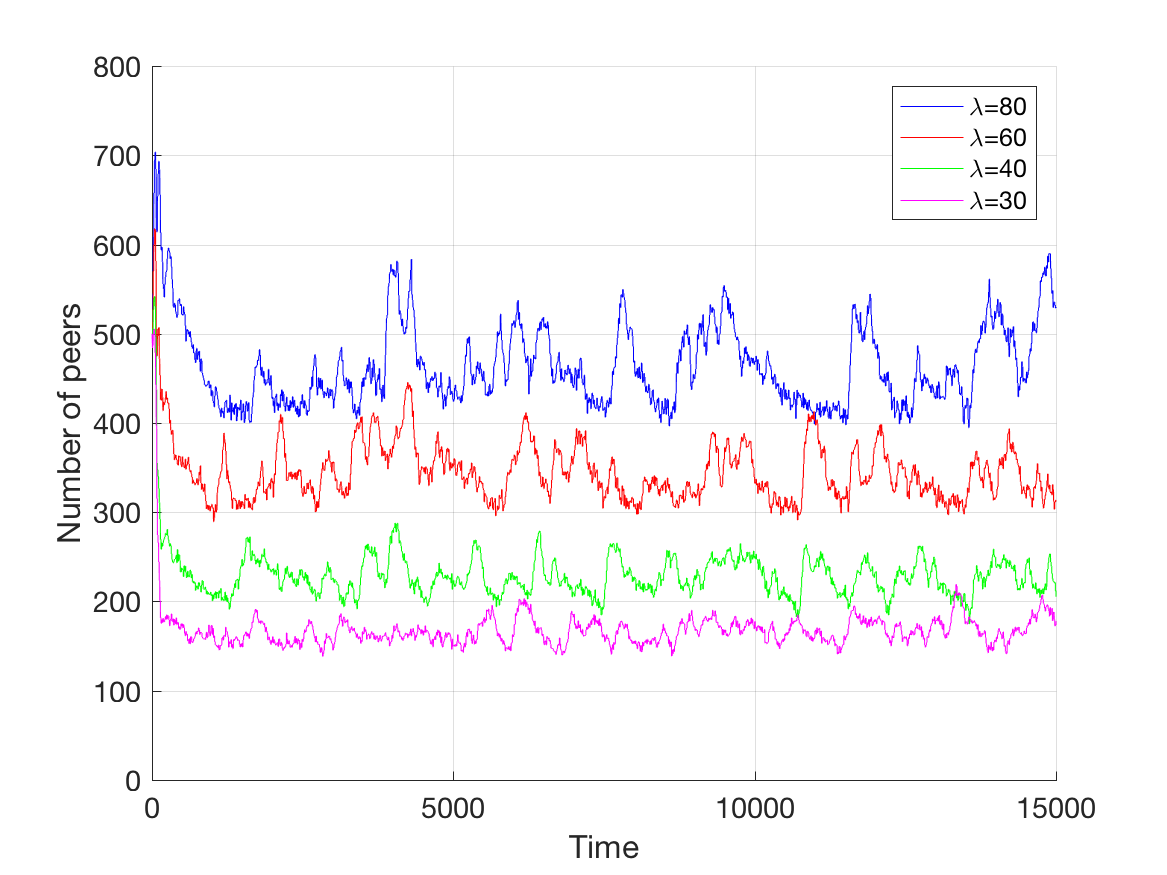}
  \captionof{figure}{The total number of peers in the network governed by the BitTorrent-like protocol with group suppression when $k=12$ and $\lambda$ is varied.}
  \label{fig:test17}
\end{minipage}\hfill
\begin{minipage}{.47\textwidth}
  \centering
  \includegraphics[width=.9\linewidth]{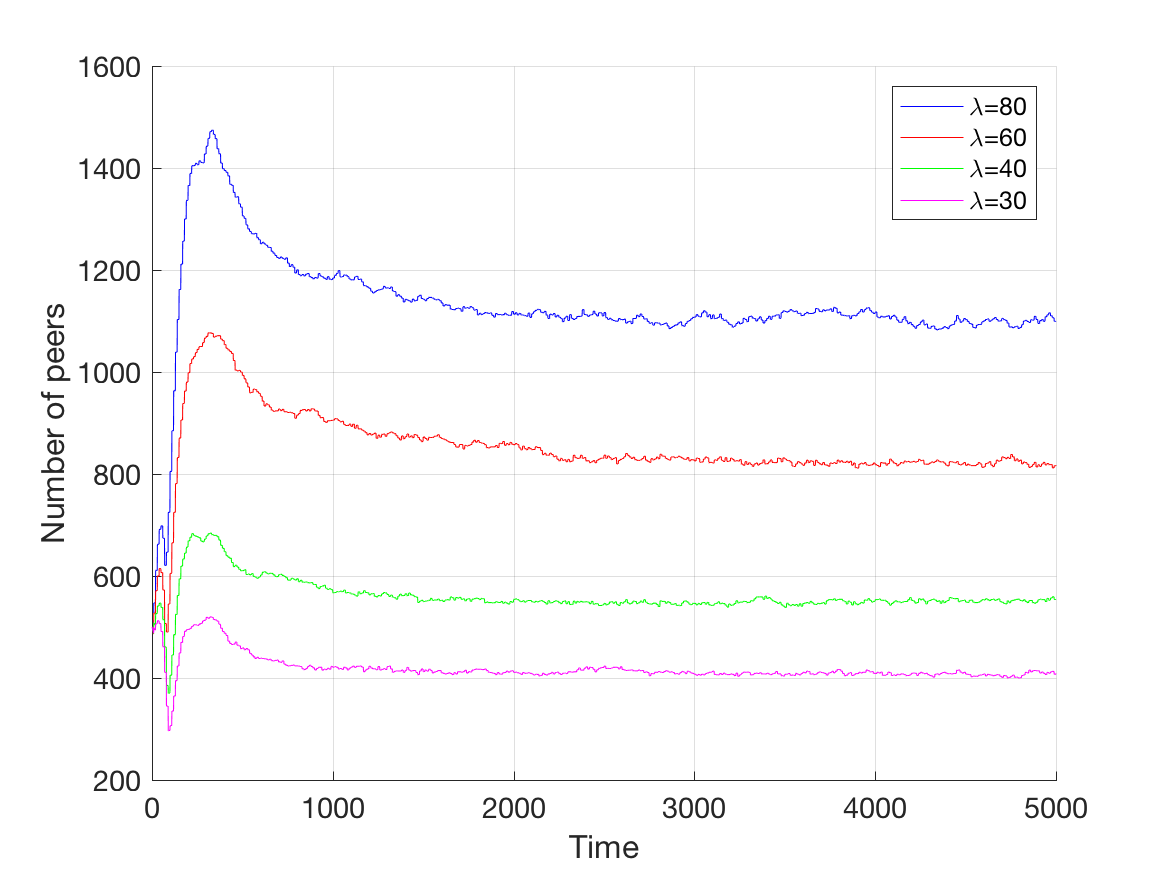}
  \captionof{figure}{The total number of peers in the network governed by BitTorrent-like protocol with group suppression when $k=48$ and $\lambda$ is varied.}
  \label{fig:test18}
\end{minipage}%
\end{figure*}

\begin{figure*}
\centering
\begin{minipage}{.47\textwidth}
  \centering
  \includegraphics[width=.9\linewidth]{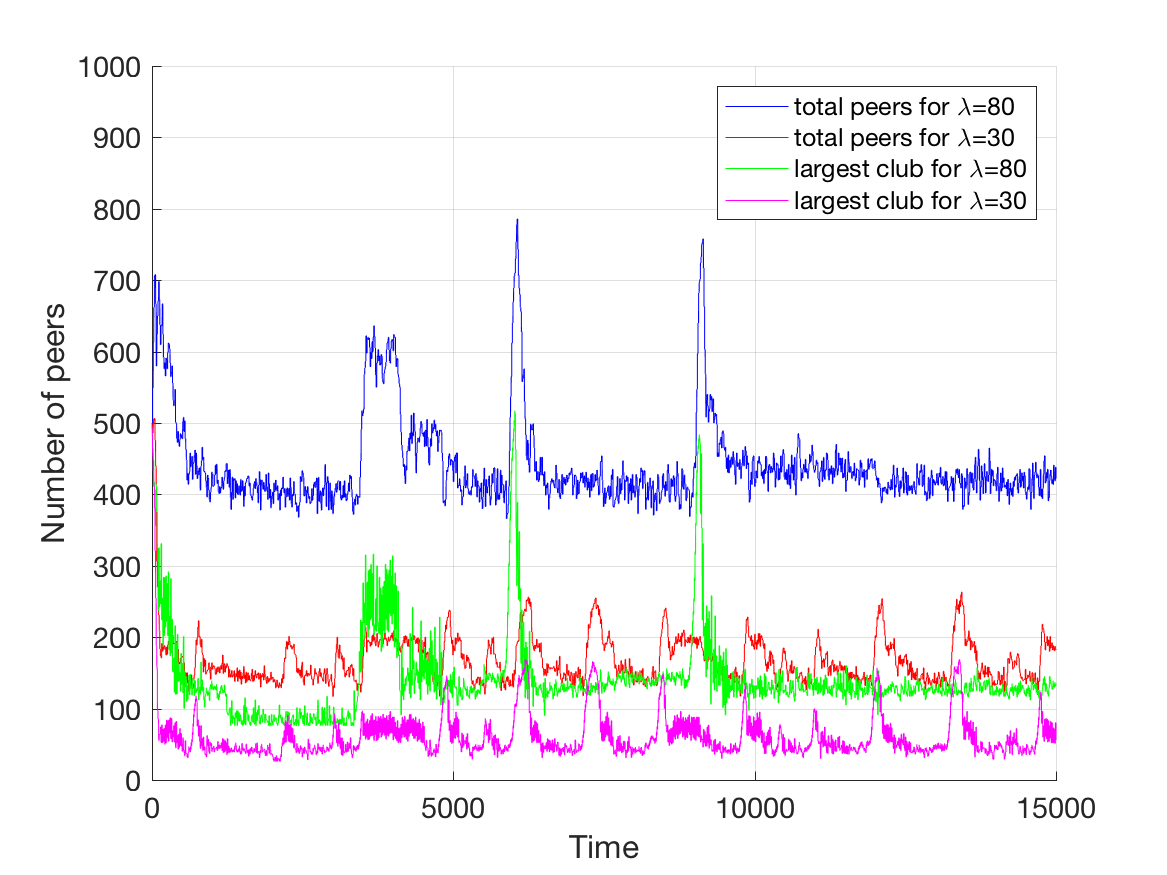}
  \captionof{figure}{In this realization of BitTorrent-like protocol with group suppression, $k=12$ is fixed. Blue and green lines show the total number of peers and the largest club when $\lambda=80$, respectively, while red and magenta lines show the total number of peers and the largest club when $\lambda=30$.}
  \label{fig:test21}
\end{minipage}\hfill
\begin{minipage}{.47\textwidth}
  \centering
  \includegraphics[width=.9\linewidth]{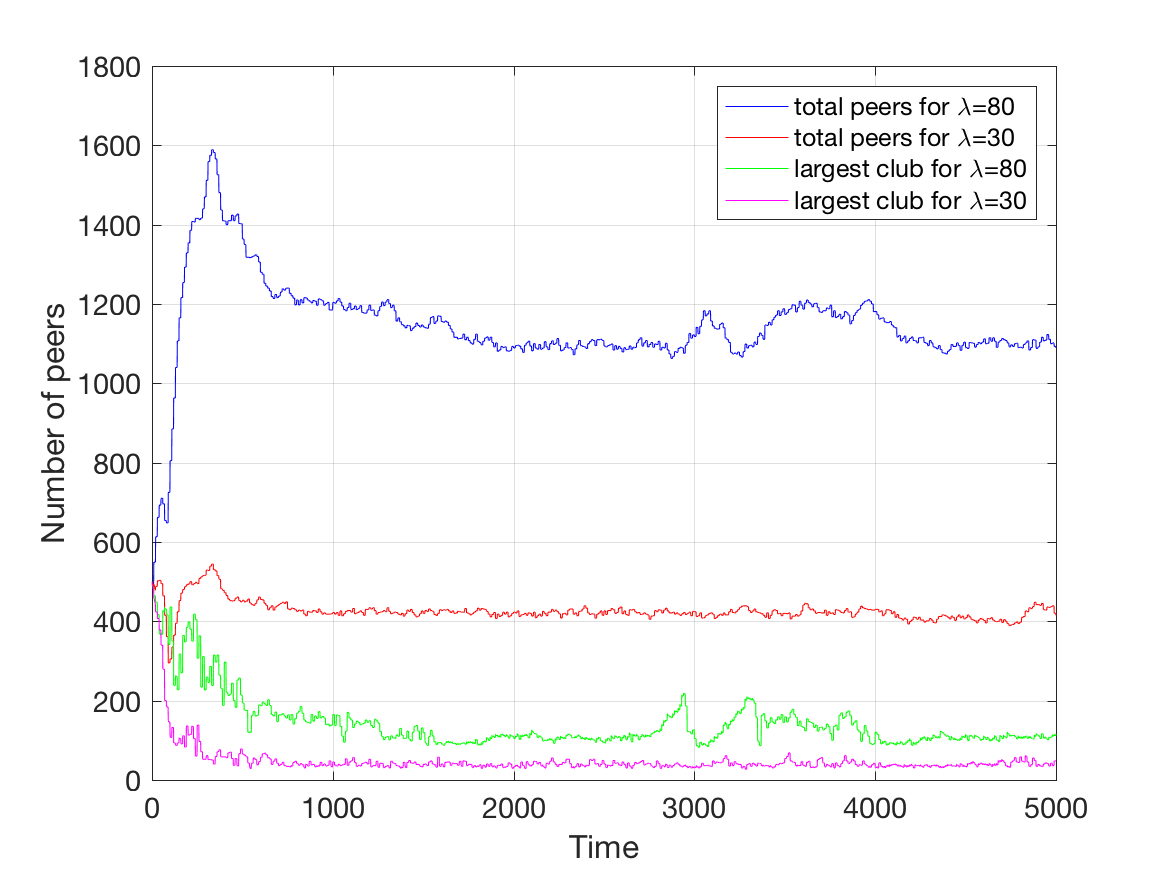}
  \captionof{figure}{In this realization of BitTorrent-like protocol with group suppression, $k=48$ is fixed. The blue and green lines show the total number of peers and the largest club when $\lambda=80$, respectively, while the red and magenta lines show the total number of peers and the largest club when $\lambda=30$.}
  \label{fig:test22}
\end{minipage}%
\end{figure*}

We now define the BitTorrent-like protocol. The original rarest first and the unchoking algorithms of the official BitTorrent protocol are preserved in the BitTorrent-like protocol although a number of simplifications were made in order to render the large scale simulations practical. First, all the upload times, arrival times and departure times are discretized and synchronous. A fixed number of peers arrives into the system every 10 seconds rather than having random individual arrivals. Also, the upload of any piece takes every incomplete peer exactly 10 seconds and they have four upload slots, meaning that they can upload at most four pieces simultaneously. Every incomplete peer's download and upload speeds are upperbounded by 40 pieces and 4 pieces per 10 seconds, respectively, in order to mimic the incomplete peer's tendency to allow much more bandwidth for the download than for upload. The seed, too, commands 4 upload slots but it finishes uploading a piece in exactly 2 seconds, which makes the seed's maximum upload speed five times an incomplete peer's maximum upload speed. In other words, the seed's maximum upload speed is 2 pieces per second and an incomplete peer's maximum upload speed is 0.4 pieces per second. Note that we do not apply strict priority, endgame mode and random first policies as our implementation operates on the level of pieces, not on the level of blocks. Out of these three features, one might suspect that endgame mode may influence stability because, after all, the missing piece syndrome results from not being able to download the last piece. 
However, endgame mode increases only the download speed of the piece but not its availability. Thus, it is reasonable to believe that not implementing endgame mode will not affect stability.

By the following modifications to some features of the original unchoking algorithm, we construct the BitTorrent-like protocol with group suppression: First, the seed ranks all interested neighbor peers in an ascending order of the number of pieces they hold. In case of a tie, the one with the larger upload ratio to the whole network is favored by the seed. It should be noted that this policy change does not add any complexity to the network as the seed already knows the piece profile of every neighbor peer. An incomplete peer first determines whether it belongs to the largest club or not based on the collection of its neighbors' piece profiles. If it does not turn out to be a member of the largest club, then the unchoking algorithm in original BitTorrent protocol proceeds. If it indeed belongs to the largest club, the incomplete peer creates the list according to the original unchoking algorithm of BitTorrent and then updates the list by removing peers that do not have more pieces than this peer. It then unchokes based on the updated list. Again, one should notice that for an incomplete peer to determine the largest club does not require any additional knowledge. Thus, the group suppression fits more naturally into BitTorrent than it does the unstructured p2p network.

The initial state of the simulation for the BitTorrent-like protocol includes one fixed seed, $494$ peers holding all the pieces but the last piece and $5$ peers possessing only the last piece; so there are $500$ peers at $t=0$. Note that these $5$ peers act as an impediment to the largest club's gravitational effect on the new peers not only by supplying the last piece to the largest club and causing them to depart, but also by giving the last piece to the new peers, which guarantees that they will not be absorbed into the largest club. They, thus, have a stabilizing effect. To show the occurrence of the missing piece syndrome for the BitTorrent-like protocol in simulations, the constant number of arrivals per $10$ second blocks is chosen from $30$, $40$, $60$ and $80$, which are equivalent to saying $\lambda$ takes on values of $3$, $4$, $6$ and $8$ arrivals per second. Note that all values of $\lambda$ are greater than the seed's maximum upload speed of $2$ pieces per second. We also remark that while Fig. \ref{fig:test21}-\ref{fig:test22} represent single realizations, all the other figures of this section demonstrate the average of five individual realization of their corresponding stochastic network. 
Fig. \ref{fig:test13} shows that the BitTorrent-like protocol falls victim to the missing piece syndrome for all four $\lambda$ values when $k$ is fixed at both $12$ and $48$. The total number of peers in Fig. \ref{fig:test13} grows roughly with rate $\lambda-U_s$ where $U_s$ is naturally interpreted as the maximum upload speed of the seed per second. 

In conclusion, the BitTorrent-like protocol can be unstable when the arrival rate of new peers overwhelms the maximum upload speed of the seed, which in turn suggests that the original BitTorrent protocol may suffer from the missing piece syndrome if all peers choose to be non-persistent. We suspect that real-life BitTorrent networks enjoys stability only because of the presence of persistent peers. In fact, another study \cite{mendes2017}, which was published concurrently with the conference version of this paper \cite{bilgen2017}, demonstrates that the missing piece syndrome unfolds in their closed BitTorrent network in which only a departure causes an arrival.

Aside from the occurrence of instability, there are two more observations worth mentioning regarding the BitTorrent-like protocol although we do not include the corresponding figures. First, increasing the number of pieces $k$ in the file seems to improve the stability for fixed upload rate of incomplete peers. But, if the upload rate of incomplete peers is made suitably large, then the increased upload capacity of the largest club would overcome the stability enhancing effect of increasing the number of pieces in the file. Second, unlike the unstructured p2p networks, increasing the arrival rate of peers beyond some point paradoxically tips the balance of the BitTorrent-like protocol into stability. This is more likely to be a specific result of discretized nature of arrivals and, therefore, this counter-intuitive observation of the BitTorrent-like network may not hold true for the continuous time arrivals that actually take place in real life BitTorrent. 

For various ($\lambda$, $k$) pairs, we provide simulation results on the BitTorrent-like protocol with group suppression in Fig. \ref{fig:test17}-\ref{fig:test22}. We modify the initial state by starting with $499$ peers that hold all the pieces but the last piece and one fixed seed. Note that this tends to decrease the stability of the system. The stable trajectories of the total number of peers exhibited in Fig. \ref{fig:test17}-\ref{fig:test22} provides experimental support for the idea that the BitTorrent-like protocol with group suppression, as opposed to the BitTorrent-like protocol, is effective against the missing piece syndrome. 
The spikes in the total number of peers manifest themselves more intensely when $k$ is smaller while they are insensitive to $\lambda$ especially when $k$ is large. Since $k$ tends to be large in actual BitTorrent networks, the spikes should not be of any concern when it comes to stability. Yet, better understanding of why such spikes are observed could shed light on how the group suppression protocol achieves stability. It can be seen from examining Fig. \ref{fig:test21} that a buildup of the largest club in the network triggers a spike of considerable magnitude in the total number of peers. Then it can be seen in the same figure that the growth of the largest club, thanks to the group suppression protocol, cannot gain further momentum and soon the largest club starts to fade away, paving the way for a steep decline in the total number of peers. Fig. \ref{fig:test22} presents further evidence of the correlation of the spikes and the buildup of the largest club. Since no buildup of the largest club takes place after $t=500$, there exists no spikes in the total number of peers in the figure. Thus, the simulations indicate that the BitTorrent-like protocol with group suppression is stable even the number of pieces is greater than two.

\section{Conclusion}
\label{sect:conclusion}
When a p2p system with uniform random peer and piece selection policies is composed of only non-persistent peers and a fixed seed, it was known that the p2p system suffers from instability if the arrival rate of new peers $\lambda$ exceeds the seed's upload rate $U_s$ \cite{HajekMissing}. Specifically, the network ends up producing a large group of peers who possess all the pieces except a common piece and the group grows roughly with rate $\lambda-U_s$. Motivated by the anticipation of p2p networks with largely non-persistent peers and the necessity of a stable protocol independent of the arrival rate of new peers and the seed's upload rate, we proposed \textit{the group suppression protocol} that achieves stability with non-persistent peers for arbitrary arrival rate of new peers and arbitrary upload rate of the seed. The group suppression protocol aims to achieve roughly uniform piece distribution over the network by cutting off the upload from the peers with the most abundant piece profile to the peers who possess smaller or equal portion of the file. It was proven to be stable using a suitable Lyapunov function for any pair of the arrival rate of new peers and the upload rate of the seed when the file is divided into two pieces. We also conjectured that the group suppression protocol's stability can be extended to any number of the pieces in the file and we supported this conjecture with simulations.

The group suppression protocol contains some centralization elements, which may limit its practicality. According to the protocol, every incomplete peer needs to determine if it belongs to the largest club. This task cannot be carried out without the knowledge of piece profiles of every other peer in the network. Second, the protocol instructs the fixed seed to contact the peers with the smallest amount of the file. This, too, requires some central knowledge. We removed these centralization requirements by devising \textit{the decentralized group suppression protocol}. In this protocol, an incomplete peer makes a decision on the largest club membership not comparing its piece profile to the entire network but to the last $3$ contacts it has initiated. As for the fixed seed, we assume every new arrival provides the fixed seed with its id, which is quite realistic assumption for practical systems. The fixed seed uploads to the latest arrival. It stores the most recent $5$ arrivals in case the newly arrived peer have already left the network before the fixed seed's Poisson upload time. The simulations showed that the protocol is stable for all $(\lambda,k)$ pairs considered. 

Hajek and Zhu's instability result \cite{HajekMissing} was proven in the context of p2p network models with uniform random peer selection policy and uniform random useful piece selection policy. Yet, most of the practical p2p protocols apply more sophisticated policies than uniform random peer selection policy and uniform random useful piece selection policy. Thus, one may suspect that the non-persistent behavior could pose no threat to the stability of some actual p2p networks that are equipped with more refined policies than uniform random selection policies. Called the BitTorrent-like protocol, a discretized BitTorrent protocol based on the official BitTorrent protocol was implemented to see if the practical p2p networks are vulnerable to instability when the peers adopt the non-persistent behavior. Extensive simulations showed that when a large fraction of the network was comprised of the one club peers at the start of simulations and the arrival rate of new peers exceeds the upload rate of the seed, the BitTorrent-like protocol with the non-persistent peers cannot prevent the one club from growing further without any bound, which leads to instability. Thus, this unstable behavior of the discretized BitTorrent raises legitimate questions on the stability of real-life BitTorrent networks if they cease to enjoy the existence of persistent peers. 

We implemented the BitTorrent-like protocol with group suppression by employing a small number of changes to the unchoking algorithm of the official BitTorrent protocol. Remarkably, the official BitTorrent protocol readily lends itself into the implementation of the group suppression protocol because it already requires every peer to know the piece profiles of all of its neighbors, which means that the largest club for a peer can be identified by the evaluation of its neighbors rather than the evaluation of every peer in the network. We ran detailed simulations of the BitTorrent-like protocol with group suppression by varying the arrival rate of new peers and the number of pieces in the file and observed that for all $(\lambda, k)$ pairs we considered, the BitTorrent-like protocol with group suppression manages to remove the one-club, which comprised the whole network apart from the seed in the beginning. And the protocol does not let any group of peers dominate the network after it removes the one club. These two properties basically point to stability.

\section*{Acknowledgment}
The authors wish to thank Ebad Ahmed for his contribution to the early stages of this project.

\bibliographystyle{IEEEtran}
\bibliography{IEEEabrv,p2pdatabase}

%\onecolumn
\section*{Appendix}
For any given pair $(U_s, \lambda)$, first pick constants $c_1, c_2, c_3, c_4$ and $p$ such that the following conditions are satisfied simultaneously:\\
\begin{eqnarray}
c_1, c_2, c_3, c_4, p > 0  \label{cond:eq1}\\ 
0<p<\frac{1}{2} \label{cond:eq2}\\ 
c_4 < c_3 \label{cond:eq3}\\
2 < c_1 \label{cond:eq4}\\ 
3c_1 < c_2 \label{cond:eq5}\\
pc_2 < 1-p \label{cond:eq6}\\
\lambda c_3^2 < [1-p(1+c_2)][U_s(1+c_2)-2\lambda] \label{cond:eq7} \\
2(c_1-1) < c_4 (c_3-c_4) \label{cond:eq8}\\
2c_3^2 < (c_1-2)^2 \label{cond:eq9}\\
\lambda c_3^2 < U_s \left(\frac{c_2-3c_1+4}{12}\right).\label{cond:eq10}
\end{eqnarray}\\
One can find such constants by first assigning some positive values to $c_1,c_3$ and $c_4$ so that the conditions \eqref{cond:eq3}, \eqref{cond:eq4}, \eqref{cond:eq8}, \eqref{cond:eq9} are met. One such choice is (32,20,10) for $(c_1,c_3,c_4)$. Set $p$ to $\frac{1}{2(1+c_2)}$. With this assignment \eqref{cond:eq2} and \eqref{cond:eq6} are satisfied. Finally pick $c_2$ large enough that \eqref{cond:eq5}, \eqref{cond:eq7} and \eqref{cond:eq10} are met.

Before we further delve into the proof of the Main Theorem, some structural observations on the potential function are made to shorten the proof.
Define $\tilde{a}$ and $\tilde{b}$ as the new $a$ and $b$ after one of the legitimate transitions occur in the system. Recalling $a=n_0+\min(n_0,n_1)+c_1(n_1-n_0)^+ -c_2n_2$, it is obvious that
\begin{eqnarray}
\tilde{a} \leq a+2+2c_1+c_2 \label{bound:1}\\
\tilde{a} \geq a-2-2c_1-c_2.  \label{bound:2}
\end{eqnarray}
The bounds are the same for $b$ as $a$ and $b$ are symmetric in $n_1$ and $n_2$. These bounds are very loose and can be easily improved but they are good enough to serve our purposes.
Observe that $V(\boldsymbol{s})$ in \eqref{potential:eqn1} is nonnegative for all $\boldsymbol{s}$. We proceed to show that $DV(\boldsymbol{s})\leq -\epsilon$ for all sufficiently large $s>0$ as this almost proves the Main Theorem on its own. Fix some small $\epsilon >0$.\\\\
\textbf{Case 1: } $n_0 \ge (1-p)s$.\\
Under this case $ n_0 \geq (1-p)s > ps$ and  $n_1 \leq ps, n_2 \leq ps$, which shows $n_0 > n_1, n_2$. Using the definition of $a$ and $b$ and \eqref{cond:eq6},
\begin{eqnarray}
a=n_0 + n_1 - c_2n_2  \geq (1-p-pc_2)s > 0 ~\text{for all $s>0$} \label{eqn:a1}
\\
b=n_0 + n_2 - c_2n_1 \geq (1-p-pc_2)s > 0  ~\text{for all $s>0$}.\label{eqn:a2}
\end{eqnarray}
By \eqref{eqn:a1}, \eqref{eqn:a2} and \eqref{bound:2} we have
\begin{eqnarray}
\tilde{a} \geq (1-p-pc_2)s-2-2c_1-c_2 \label{eqn:b1} \\
\tilde{b} \geq (1-p-pc_2)s-2-2c_1-c_2 \label{eqn:b2}.
\end{eqnarray}
Therefore $(\tilde{a})^+ = \tilde{a}$ and $(\tilde{b})^+ = \tilde{b}$ for all sufficiently large $s$. We derive $DV(\boldsymbol{s})$ for large $s$ as follows: 
\begin{align}
DV(\boldsymbol{s}) &=  \lambda\begin{bmatrix}
(2a+1)+(2b+1)+c_3(2d+c_3)
\end{bmatrix} \label{eqn:c1}\\
& ~ +\frac{U_s}{2} \begin{bmatrix}
 (1+c_2)(-2a+1+c_2)+(c_3-c_4)(-2d+c_3-c_4)
\end{bmatrix} \label{eqn:c2} \\
& ~ +\frac{U_s}{2} \begin{bmatrix}
 (1+c_2)(-2b+1+c_2)+(c_3-c_4)(-2d+c_3-c_4)
\end{bmatrix} \label{eqn:c3}\\
& ~ +\frac{\mu  n_2n_0}{s} \begin{bmatrix}
 (1+c_2)(-2a+1+c_2)+(c_3-c_4)(-2d+c_3-c_4)
\end{bmatrix} \label{eqn:c4} \\
& ~ +\frac{\mu  n_1n_0}{s} \begin{bmatrix}
 (1+c_2)(-2b+1+c_2)+(c_3-c_4)(-2d+c_3-c_4)
\end{bmatrix} \label{eqn:c5}\\
& ~ +\frac{\mu  n_1n_2}{s} \begin{bmatrix}
 c_2(2a+c_2)+ c_4(-2d+c_4)+(-2b+1)
\end{bmatrix} \label{eqn:c6}\\
& ~ +\frac{\mu  n_2n_1}{s} \begin{bmatrix}
 c_2(2b+c_2)+ c_4(-2d+c_4)+(-2a+1)
\end{bmatrix} \label{eqn:c7}.
\end{align}
Consider the term \eqref{eqn:c4}. \eqref{eqn:c4} corresponds to an upload from $type~2$ peers to $type~0$ peers. Using $\tilde{a}>0$ and $\tilde{b}>0$ for large $s$,
\begin{align*}
V(\boldsymbol{y})&= (\tilde{a}^+)^2 + (\tilde{b}^+)^2 +(\tilde{d})^2 \\
&= \tilde{a}^2 + \tilde{b}^2 +\tilde{d}^2 .
\end{align*}Thus, 
\begin{align*}
V(\boldsymbol{y})-V(\boldsymbol{x})&= (\tilde{a}^2 + \tilde{b}^2 +\tilde{d}^2)-(a^2 + b^2 +d^2) \\
&=(\tilde{a}^2-a^2)+(\tilde{b}^2-b^2)+(\tilde{d}^2-d^2)\\
&=[(a-1-c_2)^2 - a^2]+[b^2-b^2]+[(d-c_3+c_4)^2-d^2]\\
&=(1+c_2)(-2a+1+c_2)+(c_3-c_4)(-2d+c_3-c_4),
\end{align*}
and \[q(\boldsymbol{y},\boldsymbol{x})=\frac{\mu n_2n_0}{s}.\]
The other terms can be derived similarly.\\We claim that the sum of \eqref{eqn:c4} and \eqref{eqn:c6} and the sum of \eqref{eqn:c5} and \eqref{eqn:c7} are nonpositive for large $s$. To show this for sum of  \eqref{eqn:c4} and \eqref{eqn:c6}, note that for large $s$ the only positive term inside the brackets of both expressions is $c_2(2a+c_2)$. Yet the first term of \eqref{eqn:c4} eliminates this positive term for large $s$ because 
\begin{equation*}
\left[\frac{\mu n_2 n_0}{s}(1+c_2)(-2a+1+c_2)+\frac{\mu n_1 n_2}{s}c_2(2a+c_2)\right] \leq 0 ~\text{for large $s$.}
\end{equation*}One can justify omitting \eqref{eqn:c5} and \eqref{eqn:c7} in the same fashion. Omitting these terms, we are left with:
\begin{align}
DV(\boldsymbol{s}) &\leq  \lambda\begin{bmatrix}
(2a+1)+(2b+1)+c_3(2d+c_3)
\end{bmatrix} \label{eqn:d1}\\
& ~ +\frac{U_s}{2} \begin{bmatrix}
 (1+c_2)(-2a+1+c_2)+(c_3-c_4)(-2d+c_3-c_4)
\end{bmatrix} \label{eqn:d2} \\
& ~ +\frac{U_s}{2} \begin{bmatrix}
 (1+c_2)(-2b+1+c_2)+(c_3-c_4)(-2d+c_3-c_4)
\end{bmatrix}\label{eqn:d3}.
\end{align}
We drop the second terms of \eqref{eqn:d2} and \eqref{eqn:d3}, which are negative, and rearrange the inequality as follows:
\begin{align}
DV(\boldsymbol{s}) &<  a\begin{bmatrix}
2\lambda-U_s(1+c_2)
\end{bmatrix} \\
& ~ +b \begin{bmatrix}
 2\lambda-U_s(1+c_2)
\end{bmatrix} \\
& ~ +2\lambda c_3d + k,
\end{align} where $k$ is just some constant.
\eqref{cond:eq6} and \eqref{cond:eq7} imply $2\lambda-U_s(1+c_2) < 0$ and observe that $\min(a,b) \geq (1-p(1+c_2))s$ and $d \leq c_3s$. Then,
\begin{align*}
DV(\boldsymbol{s}) &<  s[1-p(1+c_2)]\begin{bmatrix}
2\lambda-U_s(1+c_2)
\end{bmatrix} \\
& ~ +s[1-p(1+c_2)] \begin{bmatrix}
 2\lambda-U_s(1+c_2)
\end{bmatrix} \\
& ~ +2\lambda c_3^2s + k.
\end{align*}
Since $[1-p(1+c_2)][U_s(1+c_2)-2\lambda]> \lambda c_3 ^2$ is met by \eqref{cond:eq7}, one can reduce the inequality to $DV(\boldsymbol{s}) < -K_1s + k$ where $K_1$ is some positive constant. Thus, $DV(\boldsymbol{s})\leq -\epsilon$ for all sufficiently large $s$.\\\\
\textbf{Case 2: } $n_0 \geq  n_1 \geq  n_2$ and $n_i < (1-p)s$ for every $i$.\\ Case 2 implies the following: \begin{eqnarray*}
&\frac{s}{3}\leq n_0 < (1-p)s \\
&\frac{sp}{2} < n_1 \leq \frac{s}{2}\\
&0 \leq n_2 \leq \frac{s}{3}.
\end{eqnarray*}
The potential function might reach boundaries and display piecewise linearity calling for careful handling of the drift. Thus, Case 2 will be broken into several subcases. \\\\
\textbf{Case 2.1: }Conditions of Case 2 and $n_0 > n_1+1$.\\ 
Under this case, $\tilde{a}$ and $\tilde{b}$ have the following forms regardless of whichever type of one-step transition occurs:
\begin{eqnarray*}
\tilde{a}=\tilde{n}_0+\tilde{n}_1-c_2\tilde{n}_2  \\
\tilde{b}=\tilde{n}_0+\tilde{n}_2-c_2\tilde{n}_1, 
\end{eqnarray*}where the tilded $n_i$'s are the new numbers of $type~i$ peers in the next state. It is important to remember that the $n_i$'s can change by 1 at most in absolute value for one state transition.
We now derive the drift:
\begin{align}
DV(\boldsymbol{s}) &\leq  \lambda \left(\begin{bmatrix}
(2a+1)  & \text{if}~ a \geq 0 \\
1 & \text{if}~ -1 \leq a < 0 \\
0 & \text{if} ~ a <-1
\end{bmatrix} + \begin{bmatrix}
 (2b+1)  & \text{if}~ b \geq 0 \\
 1 & \text{if}~ -1 \leq b < 0 \\
 0 & \text{if}~ b <-1
\end{bmatrix}  + c_3(2d+c_3) \right) \label{eqn:e1} \\
& ~ +\frac{U_s}{2} \left(\begin{bmatrix}
(1+c_2)(-2a+1+c_2)  & \text{if}~ a \geq 1+c_2 \\
0 & \text{if}~ 0 \leq a < 1+c_2 \\
0 & \text{if} ~ a < 0
\end{bmatrix} + (c_3-c_4)(-2d+c_3-c_4) \right) \label{eqn:e2} \\
& ~ +\frac{U_s}{2} \left(\begin{bmatrix}
(1+c_2)(-2b+1+c_2)  & \text{if}~ b \geq 1+c_2 \\
0 & \text{if}~ 0 \leq b < 1+c_2 \\
0 & \text{if} ~ b < 0
\end{bmatrix} + (c_3-c_4)(-2d+c_3-c_4) \right) \label{eqn:e3} \\
& ~ +\frac{\mu n_1n_0}{s} \left(\begin{bmatrix}
(1+c_2)(-2b+1+c_2)  & \text{if}~ b \geq 1+c_2 \\
0 & \text{if}~ 0 \leq b < 1+c_2 \\
0 & \text{if} ~ b < 0
\end{bmatrix} + (c_3-c_4)(-2d+c_3-c_4) \right)\label{eqn:e4}  \\
& ~ +\frac{\mu n_2n_0}{s} \left(\begin{bmatrix}
(1+c_2)(-2a+1+c_2)  & \text{if}~ a \geq 1+c_2 \\
0 & \text{if}~ 0 \leq a < 1+c_2 \\
0 & \text{if} ~ a < 0
\end{bmatrix} + (c_3-c_4)(-2d+c_3-c_4) \right) \label{eqn:e5} \\
& ~ +\frac{\mu n_1n_2}{s} \left(\begin{bmatrix}
c_2(2a+c_2)  & \text{if}~ a \geq 0 \\
c_2 ^2 & \text{if}~ -c_2 \leq a < 0 \\
0 & \text{if} ~ a < -c_2 
\end{bmatrix}+ \begin{bmatrix}
 -(2b-1)  & \text{if}~ b \geq 1 \\
 0 & \text{if}~ 0 \leq b < 1 \\
 0 & \text{if}~ b < 0
\end{bmatrix}  + c_4(-2d+c_4) \right)\label{eqn:e6} \\
& ~ +\frac{\mu n_2n_1}{s} \left(\begin{bmatrix}
c_2(2b+c_2)  & \text{if}~ b \geq 0 \\
c_2 ^2 & \text{if}~ -c_2 \leq b < 0 \\
0 & \text{if} ~ b < -c_2 
\end{bmatrix}+ \begin{bmatrix}
 -(2a-1)  & \text{if}~ a \geq 1 \\
 0 & \text{if}~ 0 \leq a < 1 \\
 0 & \text{if}~ a < 0
\end{bmatrix}  + c_4(-2d+c_4) \right)\label{eqn:e7}.
\end{align}
This inequality differs substantially from the one presented in Case 1. Indeed, the expression in Case 1 was an equality while we have inequality here. Therefore, a brief explanation is in order. Consider the first term of \eqref{eqn:e1}. \eqref{eqn:e1} corresponds to an arrival of a $type~0$ peer and the first term in parenthesis of \eqref{eqn:e1} corresponds to $(\tilde{a}^+)^2-(a^+)^2$. Observe that $\tilde{a}=a+1$ in this case. Therefore,
\begin{align*}
&(\tilde{a}^+)^2-(a^+)^2=((a+1)^+)^2-(a^+)^2=
\begin{bmatrix}
(2a+1)  & \text{if}~a\geq0\\
(a+1)^2 & \text{if}~ -1 \leq a <0\\
0 & \text{if}~ a<-1
\end{bmatrix}\\
&\leq \begin{bmatrix}
(2a+1)  & \text{if}~a\geq 0\\
1 & \text{if}~ -1 \leq a <0\\
0 & \text{if}~ a<-1
\end{bmatrix}.
\end{align*} The reader can follow the same procedure to derive the remainder of the inequality. The key part is to write $\tilde{a}$ and $\tilde{b}$ in terms of $a$ and $b$, respectively. \\
We first give analysis for $\min(a,b) \geq 1+c_2$ as this is more involved regime to consider. Break down the first term of \eqref{eqn:e4} as $-\frac{\mu n_1 n_0}{s} 2c_2b + \frac{\mu n_1 n_0}{s}(-2b+c_2^2+2c_2+1)$. Note that the positive term of \eqref{eqn:e7} is $\frac{\mu n_1n_2}{s} 2c_2b + \frac{\mu n_1n_2}{s} c_2^2$. Notice that $-\frac{\mu n_1 n_0}{s} 2c_2b +\frac{\mu n_1n_2}{s} 2c_2b \leq 0$. So we discard these two terms from the inequality to obtain another inequality. $\frac{\mu n_1n_2}{s} c_2^2$ of \eqref{eqn:e7} is put back into the parenthesis of \eqref{eqn:e7}. After removing the nonpositive middle term inside the parenthesis of \eqref{eqn:e7}, we obtain
\begin{equation*}
c_2^2+c_4(-2d+c_4) \leq c_2^2-2c_4^2s+c_4^2 <0 ~~ \text{for all sufficiently large}~s,
\end{equation*}
which allows for the removal of \eqref{eqn:e7} for large $s$. Similarly, we discard \eqref{eqn:e6}. For \eqref{eqn:e5}, its first term in parenthesis was broken down and used to eliminate \eqref{eqn:e6}'s positive $a$ term. \eqref{eqn:e5} now becomes 
\begin{align*}
&\frac{\mu n_2 n_0}{s}[(-2a+c_2^2+2c_2+1)+(c_3-c_4)(-2d+c_3-c_4)]\\
& \leq\frac{\mu n_2 n_0}{s}[(c_2^2+2c_2+1)+(c_3-c_4)(-2c_4s+c_3-c_4)]\\ 
& \leq 0 ~~\text{for large $s$.}
\end{align*}Therefore we also discard \eqref{eqn:e5}. We write the new inequality:
\begin{align*}
DV(\boldsymbol{s}) &\leq  \lambda [(2a+1)+(2b+1)+c_3(2d+c_3)]\\
& ~ +\frac{U_s}{2} [(1+c_2)(-2a+1+c_2)+(c_3-c_4)(-2d+c_3-c_4)]\\
& ~ +\frac{U_s}{2} [(1+c_2)(-2b+1+c_2)+(c_3-c_4)(-2d+c_3-c_4)\\
& ~ +\frac{\mu n_1 n_0}{s} [(-2b+c_2^2+2c_2+1)+(c_3-c_4)(-2d+c_3-c_4)].
\end{align*}
Notice that we have intentionally not discarded the rest of \eqref{eqn:e4} because it is key to proving $DV(\boldsymbol{s}) \leq \epsilon$ as can be seen in a moment. The drift can be reduced to the following for large $s$ using $n_0 \geq \frac{s}{3}$, $n_1 > \frac{sp}{2}$ and $ d > c_4s$:
\begin{equation*}
DV(\boldsymbol{s}) \leq \frac{-\mu s^2 p(c_3-c_4)c_4}{3} + k_2s+k_3 \leq -\epsilon ~~\text{for all sufficiently large}~s,
\end{equation*}where and $k_2,k_3$ are some constants.\\\\
Suppose that $a<1+c_2$ or $b<1+c_2$ or both at the same time. Then \eqref{eqn:e4} still admits a negative quadratic term. The positive terms inside the parenthesis of \eqref{eqn:e6} and \eqref{eqn:e7} are bounded by $c_2(2+3c_2)$ while the third terms inside the parenthesis of \eqref{eqn:e6} and \eqref{eqn:e7} do not exceed $-c_4^2s$. This makes \eqref{eqn:e6} and \eqref{eqn:e7} nonpositive for sufficiently large $s$ and they can be discarded. Also we may omit \eqref{eqn:e5} because it is nonpositive for large $s$. Thus $DV(\boldsymbol{s})\leq -\epsilon$ for all sufficiently large $s$.\\\\\\
\textbf{Case 2.2: } Conditions of Case 2 and $n_0 =  n_1+1$.\\
We have 
\begin{align}
b&=n_0+n_2-c_2n_1=s-n_1-c_2n_1=s-(1+c_2)n_1 \label{eqn:f1}\\
&\geq s-\left(\frac{s}{3} -1\right)(1+c_2)=s \left(1-\frac{1+c_2}{3}\right)+1+c_2. \label{eqn:f2}
\end{align}Thus, $b<0$ for all sufficiently large $s$. \eqref{eqn:f2} also implies that $\tilde{b} \leq b+2+2c_1+c_2 < 0$ for all sufficiently large $s$. So we drop all $b$ terms from the drift inequality and we obtain for large $s$,\\
\begin{align}
DV(\boldsymbol{s}) &\leq  \lambda \left(\begin{bmatrix}
(2a+1)  & \text{if}~ a \geq 0 \\
1 & \text{if}~ -1 \leq a < 0 \\
0 & \text{if} ~ a <-1
\end{bmatrix} + c_3(2d+c_3) \right) \label{eqn:g1} \\
& ~ +\frac{U_s}{2} \left(\begin{bmatrix}
(1+c_2)(-2a+1+c_2)  & \text{if}~ a \geq 1+c_2 \\
0 & \text{if}~ 0 \leq a < 1+c_2 \\
0 & \text{if} ~ a < 0
\end{bmatrix} + (c_3-c_4)(-2d+c_3-c_4) \right) \label{eqn:g2} \\
& ~ +\frac{U_s}{2} \left(\begin{bmatrix}
(c_1-1)(2a+c_1-1)  & \text{if}~ a \geq 0 \\
(c_1-1)^2 & \text{if}~ -(c_1-1) \leq a < 0 \\
0 & \text{if} ~ a < -(c_1-1)
\end{bmatrix} + (c_3-c_4)(-2d+c_3-c_4) \right) \label{eqn:g3}  \\
& ~ +\frac{\mu n_2n_0}{s} \left(\begin{bmatrix}
(1+c_2)(-2a+1+c_2)  & \text{if}~ a \geq 1+c_2 \\
0 & \text{if}~ 0 \leq a < 1+c_2 \\
0 & \text{if} ~ a < 0
\end{bmatrix} + (c_3-c_4)(-2d+c_3-c_4) \right) \label{eqn:g4}  \\
& ~ +\frac{\mu n_1n_0}{s} \left(\begin{bmatrix}
(c_1-1)(2a+c_1-1)  & \text{if}~ a \geq 0 \\
(c_1-1)^2 & \text{if}~ -(c_1-1) \leq a < 0 \\
0 & \text{if} ~ a < -(c_1-1)
\end{bmatrix} + (c_3-c_4)(-2d+c_3-c_4) \right) \label{eqn:g5}  \\
& ~ +\frac{\mu n_1n_2}{s} \left(\begin{bmatrix}
c_2(2a+c_2)  & \text{if}~ a \geq 0 \\
c_2 ^2 & \text{if}~ -c_2 \leq a < 0 \\
0 & \text{if} ~ a < -c_2 
\end{bmatrix}+ c_4(-2d+c_4) \right) \label{eqn:g6}\\
& ~ +\frac{\mu n_2n_1}{s} \left( \begin{bmatrix}
 -(2a-1)  & \text{if}~ a \geq 1 \\
 0 & \text{if}~ 0 \leq a < 1 \\
 0 & \text{if}~ a < 0
\end{bmatrix}  + c_4(-2d+c_4) \right). \label{eqn:g7}
\end{align} If $a \geq 1+c_2$, the sum of \eqref{eqn:g4} and \eqref{eqn:g6} yields nonpositive result for all sufficiently large $s$ by making use of $n_0=n_1+1$ and $d\geq c_4s$. If $a < 1+c_2$, then
the term in \eqref{eqn:g6} does not exceed $\frac{\mu n_1n_2}{s}[c_2(1+3c_2)+c_4(-2d+c_4)]\leq0$ for all sufficiently large $s$. 
So discard \eqref{eqn:g4}, \eqref{eqn:g6} and \eqref{eqn:g7} in the large $s$ regime. Note that $c_1-1 < (c_3-c_4)c_4$ is implied by \eqref{cond:eq6}. Therefore, we write using $n_0=n_1+1 \geq \frac{s}{3}$, $a=n_0+n_1-c_2n_2 \leq s$ and $d\geq c_4s$ :
\begin{equation*}
DV(\boldsymbol{s}) \leq \frac{2s^2[(c_1-1)-c_4(c_3-c_4)]}{9}+ k_1s+k_2\leq -\epsilon 
\end{equation*} for all sufficiently large $s$.\\\\
\textbf{Case 2.3: } Conditions of Case 2 and $n_0=n_1$.\\
Under these constraints on $n_0, n_1, n_2$, we have $n_0=n_1\geq \frac{s}{3}$. For $b$, 
\begin{align*}
b&=n_0+\min(n_0,n_2)+c_1(n_2-n_0)^+ -c_2n_1\\
&=n_0+n_2-c_2n_1=s-n_1-c_2n_1\leq s\left(1-\frac{1+c_2}{3}\right).
\end{align*} Since $b \leq -ks$ for some $k>0$, $\tilde{b}\leq b+2+2c_1+c_2 < 0$ for all sufficiently large $s$. Therefore all $b$ terms disappear from the drift.
\begin{align}
DV(\boldsymbol{s}) &\leq  \lambda \left(\begin{bmatrix}
(2a+1)  & \text{if}~ a \geq 0 \\
1 & \text{if}~ -1 \leq a < 0 \\
0 & \text{if} ~ a <-1
\end{bmatrix} + c_3(2d+c_3) \right) \label{eqn:h1}\\
& ~ +\frac{U_s}{2} \left(\begin{bmatrix}
(c_2-c_1+2)(-2a+c_2-c_1+2)  & \text{if}~ a \geq c_2-c_1+2 \\
0 & \text{if}~ 0 \leq a < c_2-c_1+2 \\
0 & \text{if} ~ a < 0
\end{bmatrix} + (c_3-c_4)(-2d+c_3-c_4) \right)\label{eqn:h2}  \\
& ~ +\frac{U_s}{2} \left(\begin{bmatrix}
2(c_1-1)(2a+2c_1-2)  & \text{if}~ a \geq 0 \\
(2c_1-2)^2 & \text{if}~ -2(c_1-1) \leq a < 0 \\
0 & \text{if} ~ a < -2(c_1-1)
\end{bmatrix} + (c_3-c_4)(-2d+c_3-c_4) \right)\label{eqn:h3}  \\
& ~ +\frac{\mu n_1n_0}{s}\left(\begin{bmatrix}
2(c_1-1)(2a+2c_1-2)  & \text{if}~ a \geq 0 \\
(2c_1-2)^2 & \text{if}~ -2(c_1-1) \leq a < 0 \\
0 & \text{if} ~ a < -2(c_1-1)
\end{bmatrix} + (c_3-c_4)(-2d+c_3-c_4) \right) \label{eqn:h4} \\
& ~ +\frac{\mu n_2n_0}{s} \left(\begin{bmatrix}
(c_2-c_1+2)(-2a+c_2-c_1+2)  & \text{if}~ a \geq c_2-c_1+2 \\
0 & \text{if}~ 0 \leq a < c_2-c_1+2 \\
0 & \text{if} ~ a < 0
\end{bmatrix} + (c_3-c_4)(-2d+c_3-c_4) \right)\label{eqn:h5}  \\
& ~ +\frac{\mu n_2n_1}{s} \left(\begin{bmatrix}
(-2a+1)  & \text{if}~ a \geq 1 \\
0 & \text{if}~ 0 \leq a < 1 \\
0 & \text{if} ~ a < 0 
\end{bmatrix}+ c_4(-2d+c_4) \right) \label{eqn:h6}\\
& ~ +\frac{\mu n_1n_2}{s} \left( \begin{bmatrix}
 c_2(2a+c_2)  & \text{if}~ a \geq 0 \\
 c_2^2 & \text{if}~ -c_2 \leq a < 0 \\
 0 & \text{if}~ a < 0
\end{bmatrix}  + c_4(-2d+c_4) \right).\label{eqn:h7}
\end{align}
If $a \geq c_2-c_1+2$, $\frac{\mu n_1n_2}{s}2ac_2$ of \eqref{eqn:h7} is cancelled out by $\frac{\mu n_2n_0}{s}(-2c_2a)$ of \eqref{eqn:h5}. If not, the positive term of \eqref{eqn:h7} can be upperbounded by $\frac{\mu n_2n_0}{s}c_2(2(c_2-c_1+2)+c_2)$, which is eliminated by the second term of \eqref{eqn:h7} in the large $s$ regime. After $-2c_2a$ is used to eliminate the positive term of \eqref{eqn:h7}, we would like to make sure the remainder of \eqref{eqn:h5} is negative for large $s$. Note that  
\begin{align*}
&a= n_0+n_1-c_2n_2 \leq s \\
&d=c_3n_0+c_4n_1+c_4n_2 > c_4s.
\end{align*}
Since $(c_1-2) < c_4(c_3-c_4)$ is satisfied by \eqref{cond:eq8}, the expression inside \eqref{eqn:h5} is negative for large $s$. So we omit it along with \eqref{eqn:h6} and \eqref{eqn:h7}.\\
The expression in the parenthesis of \eqref{eqn:h4} does not exceed $-ks$ for some $k>0$ for all sufficiently large $s$  by \eqref{cond:eq8} and, therefore, \eqref{eqn:h4} admits the only quadratic term in the inequality, which is negative. Thus $DV(\boldsymbol{s})\leq -\epsilon$ for all sufficiently large $s$.\\\\
\textbf{Case 3: }$n_1 \geq (1-p)s$.\\
We have 
\begin{align}
a & = n_0+\min(n_0,n_1)+c_1(n_1-n_0)^+-c_2n_2 \label{eqn:i1}\\
& =c_1n_1-[(c_1-2)n_0+c_2n_2] \label{eqn:i2}\\
&\geq c_1(1-p)s-pc_2s= s[c_1(1-p)-pc_2] \label{eqn:i3}\\
&>\frac{(c_1-2)s}{2}\label{eqn:i4}.
\end{align}\eqref{eqn:i2} is simply because of $n_1 \geq (1-p)s> ps \geq n_0$. In addition to utilizing $n_1\geq (1-p)s$, \eqref{eqn:i3} follows from the following thanks to conditions $3c_1<c_2$ and $c_1>2$:
\begin{equation*} 
\max_{\substack{n_0+n_2\leq ps \\ \min(n_0,n_2) \geq 0}}[(c_1-2)n_0+c_2n_2]=p c_2 s.
\end{equation*} 
And  the last inequality is obtained by $p c_2 < 1-p <1$ and $1-p >\frac{1}{2}$. Since $c_1>2$, then $a \geq ks$ for some $k>0$. Recalling $\tilde{a}$ is the new $a$ after one of the legitimate transitions occur, we get
\begin{equation*} 
\tilde{a} \geq a-2-2c_1-c_2 > 0 ~~\text{for all sufficiently large}~s.
\end{equation*}Therefore we conclude that both $a$ and $\tilde{a}$ are greater than any positive constant for sufficiently large $s$.
For $b$,
\begin{align*}
b &= n_0+\min(n_0,n_2)+c_1(n_2-n_0)^+-c_2n_1\\
&\leq c_1ps-c_2n_1 ~~~\text{by} ~~c_1>2\\
&\leq c_1ps-c_2(1-p)s=s[c_1p-c_2(1-p)]<0 ~~\text{for all}~~s>0.
\end{align*}Because $b\leq -ks$ for some $k>0$, we get for $\tilde{b}$,\begin{equation*} 
\tilde{b} \leq b+2+2c_1+c_2 < 0 ~~\text{for all sufficiently large}~s.
\end{equation*}That means we can drop all $b$ terms from drift inequalities that we investigate in Case 3.\\\\
\textbf{Case 3.1: } Conditions of Case 3 and $n_0 > 0$.\\
For all sufficiently large $s$,
\begin{align}
DV(\boldsymbol{s}) &=  \lambda \left[(c_1-2)(-2a+c_1-2)+ c_3(2d+c_3) \right] \label{eqn:j1} \\
& ~ +\frac{U_s}{2} \left[(2c_1-2)(2a+2c_1-2)+(c_3-c_4)(-2d+c_3-c_4) \right] \label{eqn:j2}\\
& ~ +\frac{U_s}{2} \left[
(c_2-c_1+2)(-2a+c_2-c_1+2)+(c_3-c_4)(-2d+c_3-c_4) \right]  \label{eqn:j3}\\
& ~ +\frac{\mu n_2n_0}{s} \left[
(c_2-c_1+2)(-2a+c_2-c_1+2)+(c_3-c_4)(-2d+c_3-c_4) \right] \label{eqn:j4}\\
& ~ +\frac{\mu n_2n_1}{s} \left[
c_1(-2a+c_1)+c_4(-2d+c_4)\right].\label{eqn:j5}
\end{align}Observe that \eqref{eqn:j4} and \eqref{eqn:j5} are nonpositive terms for large $s$ and we therefore omit them. \\Also 
$(2c_1-2)(2a+2c_1-2)+(c_2-c_1+2)(-2a+c_2-c_1+2)<0$ for all sufficiently large $s$ by \eqref{cond:eq5} and the fact that $a\geq ks$. Thus we may omit \eqref{eqn:j2} and \eqref{eqn:j3}. To get $DV(\boldsymbol{s})\leq -\epsilon$ for all sufficiently large $s$, we need to satisfy
$(c_1-2)2a>2c_3d$ so that $DV(\boldsymbol{s}) \leq \lambda(-ks)$ for some $k>0$. Combined with $d \leq c_3s$ and $a\geq \frac{c_1-2}{2}s$, \eqref{cond:eq9} implies $(c_1-2)2a>2c_3d$.\\\\
\textbf{Case 3.2: } Conditions of Case 3 and  $n_0 = 0$.\\
As there exists no $type~0$ peers in the system, our policy dictates the fixed seed selects a peer uniformly at random and uploads the piece that the peer needs. We have for large $s$
\begin{align}
DV(\boldsymbol{s}) &=  \lambda \left[(c_1-2)(-2a+c_1-2)+ c_3(2d+c_3) \right]  \label{eqn:k1}\\
& ~ +\frac{U_s n_1}{s} \left[c_1(-2a+c_1)+c_4(-2d+c_4) \right]  \label{eqn:k2}\\
& ~ +\frac{U_s n_2}{s} \left[
c_2(2a+c_2)+c_4(-2d+c_4) \right] \label{eqn:k3} \\
& ~ +\frac{\mu n_2n_1}{s} \left[
c_1(-2a+c_1)+c_4(-2d+c_4)\right]\label{eqn:k4}.
\end{align}
\eqref{eqn:k4} may be omitted as it is nonpositive for large $s$. Also the sum of first terms of \eqref{eqn:k2} and \eqref{eqn:k3} is nonpositive because $c_1n_1-c_2n_2 \geq c_1(1-p)s-c_2ps \geq \frac{c_1-2}{2}s$ for all sufficiently large $s$. We are left with \eqref{eqn:k1}. We prove $DV(\boldsymbol{s})\leq -\epsilon$ for all sufficiently large $s$ in the same way as in Case 3.1\\\\
\textbf{Case 4: }$n_1 \geq n_0, n_1 \geq n_2~and~n_i<(1-p)s ~~\forall i$.\\Under this case, we observe that $n_1 \geq \frac{s}{3}$ and we have for $b$,  
\begin{align*}
b &=n_0+\min(n_0,n_2)+c_1(n_2-n_0)^+ -c_2n_1\\
&\leq n_0+\min(n_0,n_2)+c_1n_2-c_2n_1\\
&\leq n_0+n_1+c_1n_2-c_2n_1\\
&\leq n_1(c_1+2-c_2)\leq \frac{s}{3}(c_1+2-c_2)<0 ~~\text{for all}~s>0.
\end{align*} Therefore, $\tilde{b} \leq b+2+2c_1+c_2 < 0$ for all sufficiently large $s$. All $b$ terms disappear in all subcases of Case 4.\\\\
\textbf{Case 4.1:} Conditions of Case 4 and $n_1 > n_0$ and $n_0>0$ and $n_1>n_2$.\\For this case, we have\\
\begin{align}
DV(\boldsymbol{s}) &\leq  \lambda \left(\begin{bmatrix}
(c_1-2)(-2a+c_1-2)  & \text{if}~ a \geq c_1-2 \\
0 & \text{if}~ 0 \leq a < c_1-2 \\
0 & \text{if} ~ a <0
\end{bmatrix} + c_3(2d+c_3) \right) \label{eqn:m1} \\
& ~ +\frac{U_s}{2} \left(\begin{bmatrix}
(c_2-c_1+2)(-2a+c_2-c_1+2)  & \text{if}~ a \geq c_2-c_1+2 \\
0 & \text{if}~ 0 \leq a < c_2-c_1+2 \\
0 & \text{if} ~ a < 0
\end{bmatrix} + (c_3-c_4)(-2d+c_3-c_4) \right) \label{eqn:m2} \\
& ~ +\frac{U_s}{2}\left(\begin{bmatrix}
(2c_1-2)(2a+2c_1-2)  & \text{if}~ a \geq 0 \\
(2c_1-2)^2 & \text{if}~ -(2c_1-2)\leq a < 0 \\
0 & \text{if} ~ a < -2(c_1-1)
\end{bmatrix} + (c_3-c_4)(-2d+c_3-c_4) \right) \label{eqn:m3} \\
& ~ +\frac{\mu n_2n_0}{s} \left(\begin{bmatrix}
(c_2-c_1+2)(-2a+c_2-c_1+2)  & \text{if}~ a \geq c_2-c_1+2 \\
0 & \text{if}~ 0 \leq a < c_2-c_1+2 \\
0 & \text{if} ~ a < 0
\end{bmatrix} + (c_3-c_4)(-2d+c_3-c_4) \right) \label{eqn:m4} \\
& ~ +\frac{\mu n_2n_1}{s} \left( \begin{bmatrix}
 c_1(-2a+c_1)  & \text{if}~ a \geq c_1 \\
 0 & \text{if}~ 0 \leq a < c_1 \\
 0 & \text{if}~ a < 0
\end{bmatrix}  + c_4(-2d+c_4) \right) \label{eqn:m5}.
\end{align}
We now investigate $DV(\boldsymbol{s})$ depending on different values of $n_2$. First, suppose $n_2<\frac{s}{2c_2}$. Then,
\begin{align}
a&=n_0(2-c_1)+c_1n_1-c_2n_2 \label{eqn:n1}\\
& \geq n_0(2-c_1)+c_1n_1-\frac{sc_2}{2c_2}\label{eqn:n2} \\
& \geq 2n_1-\frac{s}{2} \label{eqn:n3}\\
& \geq \frac{2s}{3}-\frac{s}{2}=\frac{s}{6}, \label{eqn:n4}
\end{align} where \eqref{eqn:n3} follows from $n_1\geq n_0$, $c_1>2$ and \eqref{eqn:n4} is obtained by $n_1 \geq \frac{s}{3}$.\\We omit everything except the second term in \eqref{eqn:m1} and first terms of \eqref{eqn:m2} and \eqref{eqn:m3} as the omitted terms are all negative for all sufficiently large $s$. To get $DV(\boldsymbol{s})\leq -\epsilon$, the following condition suffices:
\begin{equation*}
\lambda c_3^2 < U_s \left( \frac{c_2-3c_1+4}{12}\right).
\end{equation*}
Note that this is just \eqref{cond:eq10}.\\
If $n_2 \geq \frac{s}{2c_2}$, \eqref{eqn:m5} admits a negative quadratic term. The only remaining term that may be quadratic in $s$ is \eqref{eqn:m4}. However, the expression inside the parenthesis of \eqref{eqn:m4} is negative for all sufficiently large $s$ so it is omitted. So $DV(\boldsymbol{s}) \leq -\epsilon$ because of the negative quadratic coming from \eqref{eqn:m5}.  \\\\
\textbf{Case 4.2: } Conditions of Case 4 and $n_1>n_0$ and $n_0=0$ and $n_1>n_2$.\\
Notice that $n_1>\frac{s}{2}$ and $n_2>sp$ and the fixed seed uploads to $type~ n_1$ and $type ~n_2$ peers selected uniformly at random. We have:
\begin{align}
DV(\boldsymbol{s}) &\leq  \lambda \left(\begin{bmatrix}
(c_1-2)(-2a+c_1-2)  & \text{if}~ a \geq c_1-2 \\
0 & \text{if}~ 0 \leq a < c_1-2 \\
0 & \text{if} ~ a <0
\end{bmatrix} + c_3(2d+c_3) \right) \label{eqn:l1} \\
& ~ +\frac{U_sn_1}{s} \left(\begin{bmatrix}
c_1(-2a+c_1)  & \text{if}~ a \geq c_1 \\
0 & \text{if}~ 0 \leq a < c_1\\
0 & \text{if} ~ a < 0
\end{bmatrix} + c_4(-2d+c_4) \right) \label{eqn:l2} \\
& ~ +\frac{U_sn_2}{s}\left(\begin{bmatrix}
c_2(2a+c_2)  & \text{if}~ a \geq 0 \\
c_2 ^2 & \text{if}~ -c_2\leq a < 0 \\
0 & \text{if} ~ a < -c_2
\end{bmatrix} + c_4(-2d+c_4) \right)  \label{eqn:l3}\\
& ~ +\frac{\mu n_1n_2}{s} \left( \begin{bmatrix}
 c_1(-2a+c_1)  & \text{if}~ a \geq c_1 \\
 0 & \text{if}~ 0 \leq a < c_1 \\
 0 & \text{if}~ a < 0
\end{bmatrix}  + c_4(-2d+c_4) \right) \label{eqn:l4}.
\end{align}\
\eqref{eqn:l4}'s first term is nonpositive so we may omit it. The second term of \eqref{eqn:l4} admits negative quadratic expression because $\frac{\mu n_1n_2}{s} < \frac{\mu sp}{2}$ and $-c_42d=-2c_4^2 s$. Thus $DV(\boldsymbol{s})\leq -\epsilon$ for all sufficiently large $s$.\\\\
\textbf{Case 4.3: }Conditions of Case 4 and $n_1>n_0$ and $n_0>0$ and $n_1=n_2$.\\
Note $a=(2-c_1)n_0+c_1n_1-c_2n_2 < c_1n_1-c_2n_2=(c_1-c_2)n_1\leq(c_1-c_2)\frac{s}{3}<0$ and, therefore, $\tilde{a}\leq a+2+2c_1+c_2 <0$ for all sufficiently large $s$. All $a$'s are dropped from the drift. Recall that we dropped all $b$ terms from the drift by the analysis given in the beginning of Case 4. Thus, we get for all sufficiently large $s$:
\begin{align}
DV(\boldsymbol{s}) &=  \lambda \left(c_3(2d+c_3) \right) \label{eqn:o1}  \\
& ~ +\frac{U_s}{2} \left( (c_3-c_4)(-2d+c_3-c_4) \right)  \label{eqn:o2}  \\
& ~ +\frac{U_s}{2}\left((c_3-c_4)(-2d+c_3-c_4) \right)  \label{eqn:o3} \\
& ~ +\frac{\mu n_2n_0}{s} \left((c_3-c_4)(-2d+c_3-c_4) \right) \label{eqn:o4}  \\
& ~ +\frac{\mu n_1n_0}{s} \left((c_3-c_4)(-2d+c_3-c_4) \right) \label{eqn:o5}\\
& ~ +\frac{\mu n_2n_1}{s} \left(c_4(-2d+c_4) \right)\label{eqn:o6}\\
& ~ +\frac{\mu n_1n_2}{s} \left(c_4(-2d+c_4) \right) \label{eqn:o7}.
\end{align}
Observe that \eqref{eqn:o4} and \eqref{eqn:o5} are nonpositive for all sufficiently large $s$ and \eqref{eqn:o6}, \eqref{eqn:o7} admit negative quadratic terms. Therefore we have $DV(\boldsymbol{s})\leq -\epsilon$.\\\\
\textbf{Case 4.4: } Conditions of Case 4 and $n_1>n_0$ and $n_0=0$ and $n_1=n_2$.\\
Since $a=c_1n_1-c_2n_2=(c_1-c_2)\frac{s}{2}$, then $\tilde{a} \leq a+2+2c_1+c_2 <0$ for all sufficiently large $s$. Therefore we have:
\begin{align}
DV(\boldsymbol{s}) &=  \lambda \left(c_3(2d+c_3) \right) \label{eqn:p1} \\
& ~ +\frac{U_sn_1}{s} \left( c_4(-2d+c_4) \right) \label{eqn:p2} \\
& ~ +\frac{U_sn_2}{s}\left(c_4(-2d+c_4) \right)  \label{eqn:p3}\\
& ~ +\frac{\mu n_2n_1}{s} \left(c_4(-2d+c_4) \right) \label{eqn:p4}\\
& ~ +\frac{\mu n_1n_2}{s} \left(c_4(-2d+c_4) \right) \label{eqn:p5}.
\end{align}
First note that \eqref{eqn:p1}-\eqref{eqn:p3} do not have any quadratic term. Also \eqref{eqn:p4} and \eqref{eqn:p5} admit negative quadratic terms, which implies $DV(\boldsymbol{s})\leq -\epsilon$ for large $s$.\\\\
\textbf{Case 4.5: }Conditions of Case 4 and  $n_1=n_0 \geq n_2 $.\\
Under this case, $ \frac{s}{3} \leq n_1 \leq \frac{s}{2}$ and $n_2 \leq \frac{s}{3}$. Then $a=(2-c_1)n_0+c_1n_1-c_2n_2=2n_1-c_2n_2 \leq s$.\\
\begin{align}
DV(\boldsymbol{s}) &\leq  \lambda \left(\begin{bmatrix}
(2a+1)  & \text{if}~ a \geq 0 \\
1 & \text{if}~ -1 \leq a < 0 \\
0 & \text{if} ~ a <-1
\end{bmatrix} + c_3(2d+c_3) \right) \label{eqn:q1} \\
& ~ +\frac{U_s}{2} \left(\begin{bmatrix}
(2c_1-2)(2a+2c_1-2)  & \text{if}~ a \geq 0 \\
(2c_1-2)^2 & \text{if}~ -(2c_1-2) \leq a < 0 \\
0 & \text{if} ~ a < -(2c_1-2)
\end{bmatrix} + (c_3-c_4)(-2d+c_3-c_4) \right) \label{eqn:q2} \\
& ~ +\frac{U_s}{2} \left(\begin{bmatrix}
(c_2-c_1+2)(-2a+c_2-c_1+2)  & \text{if}~ a \geq c_2-c_1+2\\
0 & \text{if}~ 0 \leq a < c_2-c_1+2 \\
0 & \text{if} ~ a < 0
\end{bmatrix} + (c_3-c_4)(-2d+c_3-c_4) \right) \label{eqn:q3}  \\
& ~ +\frac{\mu n_1n_0}{s} \left(\begin{bmatrix}
(2c_1-2)(2a+2c_1-2)  & \text{if}~ a \geq 0 \\
(2c_1-2)^2 & \text{if}~ -(2c_1-2) \leq  a < 0 \\
0 & \text{if} ~ a < -(2c_1-2)
\end{bmatrix} + (c_3-c_4)(-2d+c_3-c_4) \right) \label{eqn:q4} \\
& ~ +\frac{\mu n_2n_0}{s} \left(\begin{bmatrix}
(c_2-c_1+2)(-2a+c_2-c_1+2)  & \text{if}~ a \geq c_2-c_1+2 \\
0 & \text{if}~ 0 \leq a < c_2-c_1+2 \\
0 & \text{if} ~ a < 0
\end{bmatrix} + (c_3-c_4)(-2d+c_3-c_4) \right) \label{eqn:q5}  \\
& ~ +\frac{\mu n_1n_2}{s} \left(\begin{bmatrix}
c_2(2a+c_2)  & \text{if}~ a \geq 0 \\
c_2 ^2 & \text{if}~ -c_2 \leq a < 0 \\
0 & \text{if} ~ a < -c_2 
\end{bmatrix}+ c_4(-2d+c_4) \right) \label{eqn:q6}\\
& ~ +\frac{\mu n_2n_1}{s} \left(\begin{bmatrix}
(-2a+1)  & \text{if}~ a \geq 1 \\
0 & \text{if}~ 0 \leq a < 1 \\
0 & \text{if} ~ a < 0 
\end{bmatrix}+ c_4(-2d+c_4) \right) \label{eqn:q7}.
\end{align}First of all, using $a\leq s$ and \eqref{cond:eq8} we get
$(2c_1-2)2a+(c_3-c_4)(-2d) \leq (2c_1-2)2s-(c_3-c_4)2c_4s \leq -ks $ for some $k>0$. This implies that \eqref{eqn:q4} admits a negative quadratic term as $n_1, n_0 \geq \frac{s}{3}$. Also omit \eqref{eqn:q7} as it is nonpositive for all sufficiently large $s$. Combining \eqref{eqn:q5} and the first term of \eqref{eqn:q6} without constants, we have
\begin{align*}
&\frac{\mu n_2 n_0}{s} [(c_2-c_1+2)(-2a)+(c_3-c_4)(-2d)]+\frac{\mu n_1n_2}{s}2ac_2\\
&= \frac{\mu n_1n_2}{s} [(c_1-2)(2a)+(c_3-c_4)(-2d)]\\
&\leq \frac{\mu n_1n_2}{s} [(c_1-2)(2s)+(c_3-c_4)(-2c_4s)] \leq \frac{\mu n_1n_2}{s} (-k_1s)~~~\text{for some}~k_1>0,
\end{align*} where we used $(c_3-c_4)c_4>2(c_1-1)$ and $c_1>2$ to get $(c_1-2)<c_4(c_3-c_4)$. Thus \eqref{eqn:q5} and \eqref{eqn:q6} are omitted as well. Since we have only one quadratic term that is negative, $DV(\boldsymbol{s})\leq -\epsilon$ for all sufficiently large $s$.\\\\
\textbf{Case 5: } $n_0\geq n_2\geq n_1$ and $n_i<(1-p)s$ for every $i$.\\
Note that Case 5 and Case 2 are symmetric.\\\\
\textbf{Case 6: } $n_2\geq (1-p)s$.\\
Note that Case 6 and Case 3 are symmetric.\\\\
\textbf{Case 7: } $n_2\geq n_0$, $n_2 \geq n_1$ and $n_i < (1-p)s~~ \forall i$.\\
Note that Case 7 is symmetric to Case 4.\\\\
By Case 1-7, we proved the existence of some $s_0(\epsilon)>0$ such that $DV(\boldsymbol{s})\leq -\epsilon$ for all $s\geq s_0(\epsilon)$. Now we let $b=$ max$\{DV(\boldsymbol{s}): s<s_0)\}+1$ and $\boldsymbol{C}=\{\boldsymbol{s}: s<s_0\}$, thus \eqref{foster:fst3} is achieved and $\boldsymbol{C}$ is a finite set as desired. Finally we appeal to $V(\boldsymbol{s}) \geq (c_4 s)^2$ to show \eqref{foster:fst2}.
\end{document}